\documentclass[conference]{IEEEtran}
\IEEEoverridecommandlockouts
\usepackage{cite}
\usepackage{amsmath,amssymb,amsfonts}
\usepackage{hyperref}
\usepackage{algorithmic}
\usepackage{graphicx}
\usepackage{subfig}
\usepackage{textcomp}
\usepackage{xcolor}
\usepackage{listings}
\usepackage{pifont}
\usepackage{array}
\usepackage{float}
\usepackage{booktabs}
\usepackage{threeparttable}
\usepackage{xcolor}
\usepackage{array}
\usepackage{textcomp}
\usepackage{stfloats}
\usepackage{url}
\usepackage{verbatim}
\usepackage{graphicx}
\usepackage[linesnumbered,lined,ruled,vlined]{algorithm2e}

\def\BibTeX{{\rm B\kern-.05em{\sc i\kern-.025em b}\kern-.08em
    T\kern-.1667em\lower.7ex\hbox{E}\kern-.125emX}}
\begin{document}

\title{PALM: A Efficient Performance Simulator for Tiled Accelerators with Large-scale Model Training }

\author{\IEEEauthorblockN{Jiahao Fang\IEEEauthorrefmark{1}, Huizheng Wang\IEEEauthorrefmark{1}, Qize Yang\IEEEauthorrefmark{1}, Dehao Kong\IEEEauthorrefmark{2}, Xu Dai\IEEEauthorrefmark{2}, Jinyi Deng\IEEEauthorrefmark{1}, Yang Hu\IEEEauthorrefmark{1}, and Shouyi Yin\IEEEauthorrefmark{1}}
\IEEEauthorblockA{\IEEEauthorrefmark{1}\textit{School of Integrated Circuits, Tsinghua University, Beijing, China }\\
\IEEEauthorrefmark{2}\textit{ Shanghai Artificial Intelligence Laboratory, Shanghai, China  } \\
wanghz22@mails.tsinghua.edu.cn }
}



\maketitle
  
\begin{abstract}
Deep learning (DL) models are piquing high interest and scaling at an unprecedented rate. To this end, a handful of tiled accelerators have been proposed to support such large-scale training tasks. However, these accelerators often incorporate numerous cores or tiles even extending to wafer-scale, substantial on-chip bandwidth, and distributed memory systems. This results in an exceedingly complex design space. Moreover, conducting actual training experiments to find optimal configurations is impractical due to time constraints. Hence, predicting the optimal mapping of various parallelisms to such tiled system architectures becomes crucial. In this study, leveraging an analysis of existing mainstream DL model training strategies, we introduce a performance simulator named PALM. PALM targets both the training and inference processes for tiled accelerators, aiming to inspire the design of current and future accelerators. Specifically, (i) we establish a scheduling mechanism among tiled accelerators based on an event-driven framework; (ii) we support user-configurable pipeline, tensor, and data parallelism on tiled accelerators, determining the absolute performance throughput under these parallelism strategies; (iii) we model the interaction of on-chip SRAM, NoC, and off-chip DRAM during operator execution.
This work is available here: https://github.com/fangjh21/PALM.
\end{abstract}


\section{Introduction}\label{sec:Intro}
Deep learning (DL) and deep neural networks (DNN) play a crucial role in advancing artificial intelligence (AI) across diverse application domains, including image processing~\cite{he2016deep,carion2020end,dosovitskiy2020image,liu2022swin,liu2021swin,rombach2022high}, natural language processing~\cite{radford2018improving,raffel2020exploring,li2022blip}, and autonomous driving~\cite{grigorescu2020survey,qi2017pointnet,zhou2018voxelnet}. As the popularity and applications of AI continue to grow, researchers are actively working to enhance the capabilities and accuracy of DNN. This involves designing more complex networks and training them with extensive datasets, often comprising millions or even billions of samples~\cite{touvron2023llama,radford2019language,brown2020language}. However, these advancements come with the challenge of extended training times and skyrocketing memory requirements, thereby fueling the need for scalable high-performance training platforms. For example, training GPT-3 (175\,B) on Nvidia Tesla V100 GPUs acquires 3.1 million hours and would cost around $\$$4.6 million~\cite{wang2022lightseq2}. Even worse, the overall size of these huge models surpasses the physical memory capacity of a single accelerator. This holds true even for contemporary GPUs equipped with substantial memory, such as the 80GB Nvidia H100 cards~\cite{choquette2022nvidia}. Therefore, numerous efforts have been devoted to expediting the training process by distributing it across multiple accelerators.

The fundamental concept behind distributed training is to allocate the independent computations of the model across multiple accelerators, facilitating parallel execution. Various parallelization strategies are available~\cite{li2020pytorch,huang2019gpipe,shoeybi2019megatron}, each with its own set of advantages and drawbacks. Identifying the appropriate type and degree of parallelism to be leveraged under different constraints (such as budget, time, memory, and ease of implementation) can significantly enhance training throughput. However, it is impractical to find the optimal type and degree of parallelism by performing actual training experiments given some specific constraints due to the prohibitive expense. Although most academic projects leverage cloud frameworks like Microsoft Azure, Google Cloud Computing, or Amazon Web Services for training their proposed models, conducting these long-running experiments on cloud-hosted systems is also expensive as users are billed per hour. Therefore, an effective prediction for the training time under given workloads, parallelism configurations, and accelerator architectures becomes an indispensable part of the distributed training system design. 


Recently, tiled accelerators~\cite{apple,gao2017tetris,gao2019tangram,jouppi2021ten,shao2019simba,wechsler2019spring} have been recognized for significant potential in DL distributed training tasks due to their higher utilization and energy efficiency~\cite{moon2021evaluating}. These accelerators feature spatial multi-tiled architectures, with each hardware tile comprising a processing element (PE) array and a global buffer, interconnected by a network on chip (NoC). Therefore, it becomes crucial to perform simulation modeling for tiled accelerators. However, existing simulators often lack DL training support on tiled accelerators for the following reasons: (i) Current simulators adopt cycle-accurate or event-driven approaches, lacking of a \textbf{scheduling mechanism to model a large number of tiles}. (ii) These simulators lack \textbf{user-configurable parallelism strategies}, ignoring users' needs to optimize performance with hybrid parallelism strategies. (iii) Tiled accelerators exhibit spatial properties that involve \textbf{interaction between DRAM and NoC bandwidth}, posing a challenge for existing analytical models to capture, while cycle-accurate models are cumbersome.

Given these insights, PALM is introduced as a simulator tailored for DL training on tiled accelerators. PALM utilizes three internal mechanisms to tackle these issues:  \textbf{(i) Virtual Tile Aggregation}, with which pipeline execution and layer-wise execution for the training of DL models ranging from tens to thousands of tiles can be modeled ; \textbf{(ii) Adaptive Parallelism Interface} which supports parallelism strategies and spatial mapping configured by users, providing them with a broad search space; \textbf{(iii) Detailed Bandwidth Model} which supports modeling bandwidth contention phenomenon on multi communication and access task. The main contributions of this work are summarized as follows:
\begin{itemize}
    \item To the best of the author's knowledge, PALM is the first simulator considering the spatial property of tiled accelerators on DL training tasks with event-driven mechanism.
    \item We identify three major challenges in modeling tiled accelerators: software overhead in simulating a large number of tiles, lack of user interfaces for configuring parallelism strategies, and difficulty in modeling  influence between DRAM and NoC with existing methods.
    \item In response to these modeling challenges, we propose three corresponding mechanisms: Virtual Tile Aggregation, Adaptive Parallelism Interface, and Detailed Bandwidth Model.
    \item Through several case studies, we demonstrate PALM's modeling accuracy. Compared to published data, our average error remains within 17\%. Additionally, we show that subtle differences in spatial mapping and parallelism within tiled accelerators result in a performance gap 2$\times$  larger. Finally, we delve into the optimization of communication across tile groups.
\end{itemize}

\section{Background}\label{sec:Background}

\begin{figure}[t]
  \centering
  \includegraphics[width=0.999\linewidth]{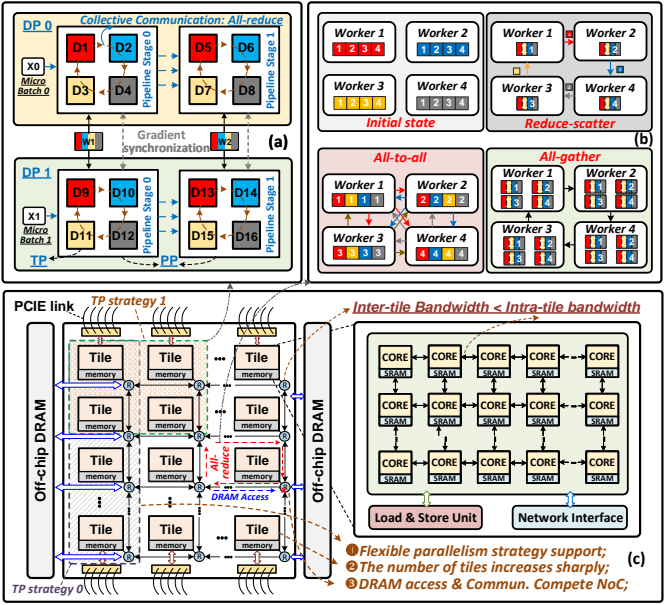}
  \caption{Diverse parallelism strategies, collective communication patterns, and typical architecture of tiled accelerators.}
  \label{fig:MTA}
\end{figure}

\subsection{Parallelism Schemes of Distributed Training}\label{subsec:DST}
\subsubsection{Data Parallelism (DP)}\label{subsubsec:DP} As shown in Fig.~\ref{fig:MTA}(a), DP means each worker utilizes the same model to train on distinct micro-batches of data~\cite{shoeybi2019megatron}. In DP, there is no synchronization between workers during forward computation, as each worker possesses a complete copy of the model. The storage for holistic structure and parameters also leads to an expensive memory footprint. Despite the elimination of data synchronization during the forward process, gradient all-reduce becomes essential as a collective operation during the backward process.

\subsubsection{Tensor Parallelism (TP)}\label{subsubsec:TP} In TP, the model weights are divided (depicted by diverse colors in Fig.~\ref{fig:MTA}(a)), while training data is duplicated across workers~\cite{narayanan2021efficient}. Consequently, each worker observes the same data but computes only a portion of the activation. The communication of these partial results is necessary across workers in layers during both forward and backward propagation. Compared to the DP, the communication cost from TP is higher, but it can effectively relieve the memory capacity pressure\,\cite{shazeer2018mesh}. This allows multiple devices to jointly serve a larger model, addressing the challenge of fitting huge models onto limited hardware resources. 

\subsubsection{Pipeline Parallelism (PP)}\label{subsubsec:PP} This parallelism entails the division of the layers of DL model among workers~\cite{huang2019gpipe}, as illustrated by the four white boxes in Fig.~\ref{fig:MTA}(a). Activations from a specific set of layers, assigned to one worker, are transferred to the subsequent set of layers, assigned to another worker. These consecutive layers operate on distinct data concurrently when the input batch is segmented into micro-batches that can be sequentially fed to the pipeline workers. However, this strategy may introduce pipeline bubbles~\cite{narayanan2019pipedream,kosson2021pipelined} or periods during which an accelerator remains idle, awaiting data from the preceding accelerator in the pipeline.

\subsection{Collective Communication}\label{subsec:CC} 
Based on the chosen parallelization strategy, models and input batches are distributed across workers. This makes communication and synchronization of data, like forward activation or weight/input gradients, among devices inevitable~\cite{rashidi2022themis}. This traffic is typically formulated and processed through collective communications. Four primary collective communication operations are key contributors in DNN training~\cite{klenk2020network,rashidi2020astra}: (i) reduce-scatter, (ii) all-gather, (iii) all-reduce, (iv) all-to-all. In Fig.~\ref{fig:MTA}(b), reduce-scatter operation sums all initial data in workers, resulting in each worker holding a portion of globally reduced data. The all-gather operation gathers the data initially distributed across workers, ensuring each worker possesses the complete data. All-reduce can be regarded as a combination of reduce-scatter followed by an all-gather operation. In the all-to-all pattern, each node is required to send a distinct portion of data to other nodes. 

\subsection {Tiled Accelerator}\label{subsec:MTA}
Fig.~\ref{fig:MTA}(c) illustrates the architecture for a tiled accelerator, which usually consists of multiple independent operating tiles. Each tile has its unique instruction queue, local memory and progresses at its own pace, which thus allows the tiled accelerators to specialize in supporting flexible dataflow and mapping. Moreover, the NoC is employed for transferring data among the tiles and synchronizing tiles at different stages throughout the program execution. Also, the NoC establishes connections among all tiles, as well as off-chip communication and memory controller blocks. As a result, each tile has access to the off-chip memory or other chips. Compared to traditional monolithic chips and single-tile SIMD GPUs, such architectures usually exhibit higher execution efficiency.  Such improved efficiency comes from employing optimized dataflow strategies to spatially/temporally partition data across the tiles and fine-grained scheduling.

\subsection {Modeling Method for DL training on Hardware}\label{subsec:SIM}       
\subsubsection{Analytical Model and Prediction Model}
The analytical model\cite{moolchandani2023amped,geoffrey2021habitat,james2020ispd} examines the DL model training process, using approximate methods to derive formulas for DL model and hardware parameters to estimate latency or energy consumption. While providing a quick assessment, its reliability is moderate and may not fully capture the dynamic features of hardware systems. The prediction model\cite{geoffrey2021habitat} gathers throughput data and hardware-related information from DL training, utilizing models like Multilayer Perceptrons (MLP) for training. However, its applicability is limited, relying heavily on specific datasets and training conditions.
\subsubsection{Simulator}
Existing simulators fall into two main categories: cycle-accurate and discrete event-driven\cite{rashidi2020astra,won2023astra}. The former delves into low-level hardware logic, processing operations within each clock cycle with fine granularity and high-precision modeling, suitable for scenarios with well-defined hardware architectures. However, drawbacks include a longer development cycle and extended software runtime. In contrast, discrete event-driven simulators' trigger changes through events, maintaining an event queue for each hardware component. These simulators demonstrate faster speeds and are ideal for early-stage hardware development and architectural exploration.
\begin{figure*}
  \centering
\centerline{\includegraphics[width=0.95\textwidth]{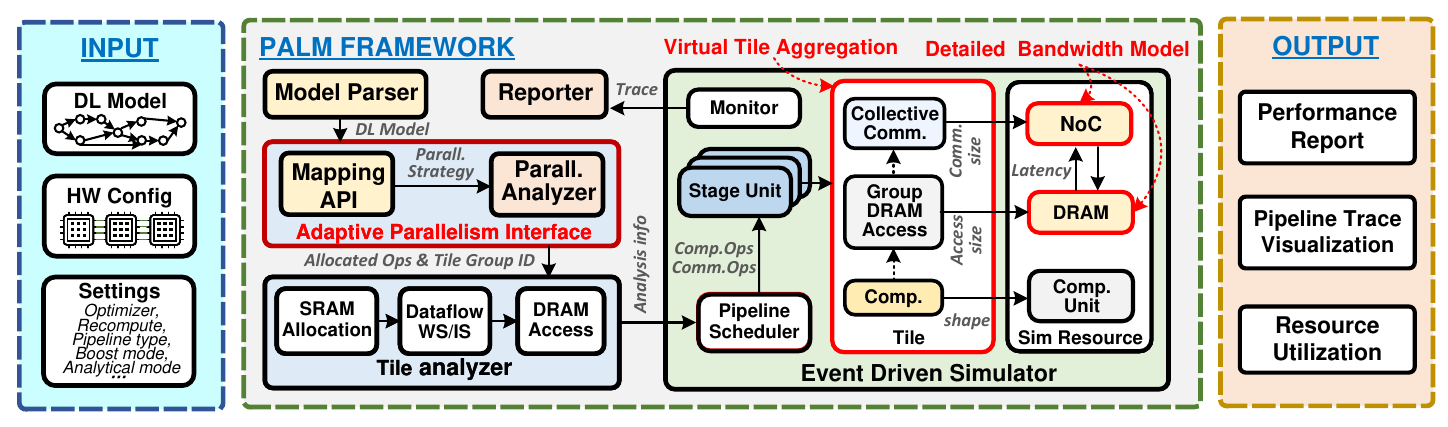}}
  \caption{An overview of PALM framework. The highlighted portions in red boxes are focal points of the work.}
  \label{fig:PALM}
\end{figure*}

\section{Motivation}
Existing simulators and analytical or prediction models primarily focus on modeling GPU clusters but lack robust support for tiled accelerators. To inspire the design of tiled accelerators for DL training, based on the property of DL models and architecture, we identify the following three essential requirements: \textbf{(i) Scheduling mechanism to model a large number of tiles; (ii) User-configurable parallelism strategies; (iii) Interaction between DRAM and NoC bandwidth.}
\subsection{scheduling mechanism to model a large number of tiles}
A sensible modeling approach is essential for simulating the training process of DL models on a substantial number of tiles, as depicted in Fig.~\ref{fig:MTA}(c). Real tiled accelerator systems exhibit a range of scales, from 4$\times$4 and 10$\times$12 \cite{cai2023inter,vasiljevic2021compute} to a wafer-scale architecture of 633$\times$633~\cite{CS}. A straightforward but very coarse approach is to assign each tile an independent thread or event queue. However, handling a large number of tiles using such a simulation mechanism would lead to a notable increase in software overhead.
Therefore, to efficiently implement a tiled accelerator simulator for DL training tasks, it is imperative to introduce a unique scheduling mechanism among tiled accelerators.
\subsection{User-configurable parallelism strategies}
Current simulators lack interfaces that support arbitrary parallelism strategies. Typically, users need to extract computation graphs with embedded parallelism information from established DL frameworks such as PyTorch and TensorFlow. This limitation prevents the direct iteration of parallelism strategies based on simulation results. Additionally, existing simulators lack support for various types of PP which is an important parallelism strategy of LLM, nor have they discussed the differences in bubble and capacity requirements under PP. In fact, the proposal of PP is mainly aimed at solving the storage problem of LLM, which has problems in resource utilization. The advantage of PP on tiled accelerators is that it fits the characteristics of a large number of tiles, can more evenly split the pipeline, increase the number of pipeline stages, and reduce the bubble ratio.
TP and DP are two inherent parallelism strategies. In the tiled accelerators, when some tiles/cores form a tile group to execute the same operator, certain dimensions must be segmented as illustrated in Fig.~\ref{fig:MTA}(a).
Hence, it is crucial to offer a flexible user-visible interface that supports parallelism across various dimensions.
\subsection{Interaction between DRAM and NoC bandwidth}
SRAM, being faster but costlier than DRAM, is utilized to temporarily store data for computation and exchange data with DRAM. Table~\ref{gpuvs} indicates that the SRAM capacity per computing power unit in tiled accelerators surpasses that in traditional GPUs. Specifically, WSE's SRAM capacity per computing unit is nearly $26\times$ that of GPU A100. Studies~\cite{kosson2021pipelined,vasiljevic2021compute} explore using SRAM to statically store frequently read data, accelerating tile computation based on dataflow. Recognizing the significant role of SRAM in computation, memory access, and communication is thus reasonable.

Efficient model training relies on DRAM with large capacity and high bandwidth. DRAM is crucial for storing extensive model parameters, intermediate activations, and optimizer states during training. Tiled accelerators, designed for high-density computing power, differ significantly from GPUs in their memory hierarchy. For example, in the WSE-2 system~\cite{CS}, of which the computing power is equivalent to 46 GPUs, there is no on-wafer DRAM; instead, DRAM is located off the wafer. Consequently, DRAM access in tiled accelerators becomes costly due to NoC routing, as depicted in Fig.~\ref{fig:MTA}(c). Therefore, modeling DRAM behavior is crucial to accurately reflect practical behaviors of tiled accelerators.

NoC acts as a physical bridge among tiles~\cite{cai2023inter,vasiljevic2021compute,ignjatovic2022wormhole}, impacting communication between pipeline stages generated by mapping and parallelism, as well as intra-stage communication. Frequent DRAM access will occupy NoC bandwidth. In Table~\ref{gpuvs}, various tiled accelerators exhibit different NoC hop counts to DRAM, presenting a disadvantage for on-chip access tasks. Additionally, in the same table, the Link bandwidth-to-DRAM bandwidth ratio is higher in tiled accelerators, providing an advantage for communication tasks.

In summary, it is essential to model the behavior of SRAM, DRAM, and NoC during the training process to accurately reflect the architectural characteristics of tiled accelerators. 

\begin{table}[t]\renewcommand{\arraystretch}{1.1}
\begin{threeparttable}
\caption{GPU VS Tiled Accelerator Hardware Parameters.}\label{gpuvs}
\begin{center}
\begin{tabular}{c|c|c|c}
\specialrule{0.12em}{0.1pt}{0.2pt}
\textbf{Hardware} &\textbf{S\_Cap.~/Comp.$^{1}$}& \textbf{D\_Hops}$^{2}$& \textbf{L\_BW~/D\_BW}$^{3}$ \\
\hline
H100\cite{choquette2022nvidia}& 0.050 & -& 0.179 \\
A100\cite{A100}&  0.128& - & 0.310 \\
\hline
Grayskull\cite{vasiljevic2021compute}  &  1.304 & 5 & 1.900 \\
Dojo D1 ~\cite{dojo}   & 1.215 & ~25 & 2.275\\
WSE2\cite{cerebras}    &  2.666 & ~316  & 1.375 \\
\specialrule{0.12em}{0.5pt}{0.6pt}
\end{tabular}
    \begin{tablenotes}
      \item[1] SRAM capacity-to-compute ratio (MB/TFLOPs@FP16); 
      \item[2] Maximum hop counts to DRAM;
      \item[3] Link bandwidth-to-DRAM bandwidth ratio. 
    \end{tablenotes}
\end{center}
\end{threeparttable}
\end{table}

\begin{table}[t]\renewcommand{\arraystretch}{1.1}
\caption{Factors affecting performance considered by PALM.}
\begin{center}
\begin{tabular}{l|c|c}
\specialrule{0.12em}{0.1pt}{0.2pt}
\textbf{Factors} &\textbf{Tpye}& \textbf{Direct affect}  \\
\hline
Pipe schedule & GPipe, (interleaved)1F1B\cite{shoeybi2019megatron}& bubble \& mem  \\
Parallelism& PP, DP, TP &  latency \&  mem.   \\
Tile dataflow\cite{samajdar2018scale} & IS, WS & access times  \\
Optimizer\cite{opt}   & SGD, Adam& mem  \\
ZERO\cite{rajbhandari2020zero}    & ZERO& latency \&  mem. \\
Congestion & NoC, DRAM  & latency   \\
\specialrule{0.12em}{0.5pt}{0.6pt}
\end{tabular}
\end{center}
\label{dims}
\end{table}

\section{The Making of PALM}
Fig.~\ref{fig:PALM} shows the overall framework of PALM and the main factors considered by PALM are concluded in Table.~\ref{dims}. The PALM is built based on the discrete event-driven framework--SimPy~\cite{simpy}. Moreover, PALM models a two-level tiled accelerator, as shown in Fig.~\ref{fig:MTA}. 
This section will introduce how to efficiently obtain performance throughput from DL models, hardware configurations, and other settings.
\begin{figure}[t]
\centering
\includegraphics[width=0.95\linewidth]{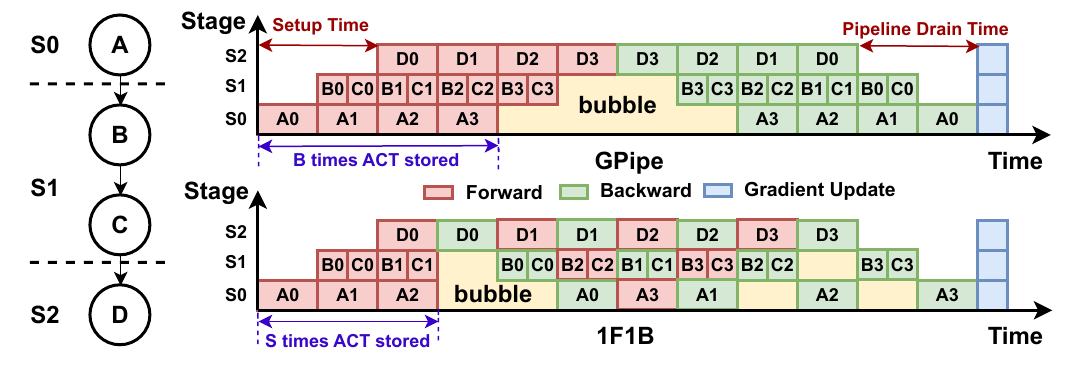}
\caption{Left: partitioning DL computation graph into pipeline stages;~Right: two pipeline scheduling methods. S means the number of stages and B means the batch size. The overhead of setup time and drain time is needed to be considered.}
\label{fig:schedule1}
\end{figure}
\begin{figure}[t]
\centering
\includegraphics[width=0.9\linewidth]{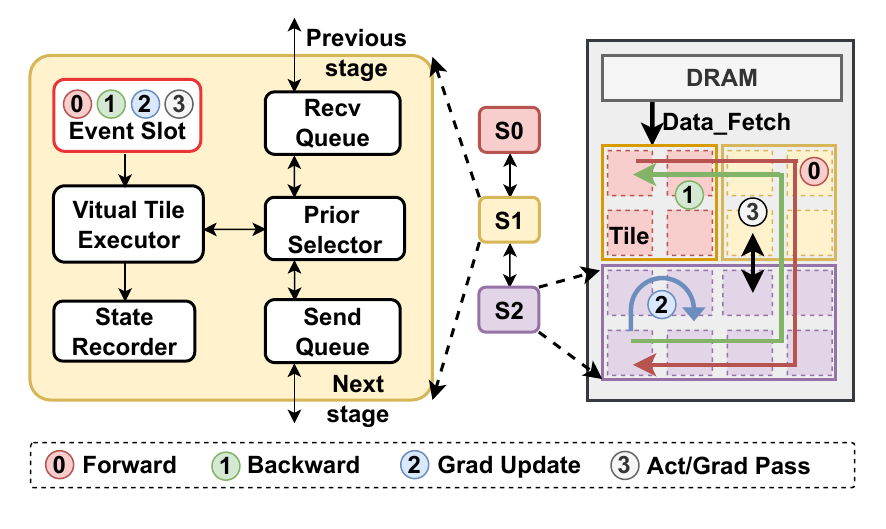}
\caption{The details of pipeline scheduler. Adjacent Stage units share a message queue.}
\label{fig:schedule2}
\end{figure}

\subsection{Virtual Tile Aggregation}
We distinguish the concept between pipeline scheduling mechanism and pipeline parallelism. The former concerns modeling the training process effectively, while the latter involves partitioning the computation graph into stages, as discussed in the next subsection.
The pipeline scheduling includes two mechanisms: pipeline execution and layer-wise execution~\cite{kosson2021pipelined}. In our modeling, layer-wise execution is treated as pipeline execution with a depth of 1. Fig.~\ref{fig:schedule2} illustrates the pipeline scheduling process: the computation graph is partitioned into stages (\verb|S0|, \verb|S1|, \verb|S2|), and each stage is mapped to a tile group based on the parallelism strategy. The pipeline is divided into three processes: \verb|Forward (FD)|  representing the forward computation of all operators in each stage, \verb|Backward (BD)|  representing the backward propagation of all operators in each stage, and  \verb|Gradient Update (GU)|  representing the gradient update process. Additionally, PALM defines \verb|Act/Grad Pass| to transfer activations/gradients across stages, serving as the start signal for the next stage.
In DL, a batch (mini-batch) is taken as the period for gradient updates. To reduce the pipeline bubble ratio, a batch is divided into multiple micro-batches, with one micro-batch executing \verb|FD| and \verb|BD|. Once all micro-batches are completed, \verb|GU| is executed. \verb|Data_Fetch| simulates the input data fetching of one micro-batch, representing the start of the first stage \verb|S0|.  
 In our scheduling mechanism, GPipe\cite{huang2019gpipe} and 1F1B\cite{shoeybi2019megatron} scheduling in Fig.~\ref{fig:schedule1} are supported . PALM places one of the four types of events into the \verb|Virtual Tile Executor| based on the signal selected by \verb|Prior Selector|. For example, in the 1F1B pipeline, priority is accorded to the execution of BD over FD. The \verb|Act/Grad Pass| between different stages is accomplished through communication events on NoC. 
 This process is primarily determined by the dependency relationships between adjacent operators in the different stages, which will be discussed in the next subsection.

Within each stage, operators are executed in the order of their dependency relationships such as op $B$ and $C$ of \verb|S1| in Fig.~\ref{fig:schedule1}, as layer-wise execution does. Operators without dependencies are executed in the pre-order rule in the computation graph or in parallel. When tiles/cores execute the same operator, they are called a \textbf{tile group}. In the tile analysis level (tile analyzer in Fig.~\ref{fig:PALM}), PALM assumes different tiles in each tile group have the same computation and memory access cost. Therefore, each stage exclusively furnishes one or a few simulated tiles representing these tiles in tile group, denoted as virtual tiles. We have coined this modeling method as \textbf{\textit{Virtual Tile Aggregation}}.

We assume that a single tile mainly consists of two entities: the tile internal logic unit and NoC router which have their own event queue. Additionally, we suppose the number of tiles is $N \times N$, and the number of stages $S$ is less than or equal to the number of layers $M$ in the computation graph. The naive modeling complexity is $\mathcal{O}(2N^2)$ for all tiles, while PALM with virtual tile aggregation reduces it to $\mathcal{O}(N^2+M)$. By incorporating an analytical model for the NoC, the complexity is further reduced to $\mathcal{O}(M)$. Given that $M$ typically falls in the range of tens to hundreds, this significantly alleviates the modeling overhead.
\begin{figure}[t]
  \centering
  \includegraphics[width=1\linewidth]{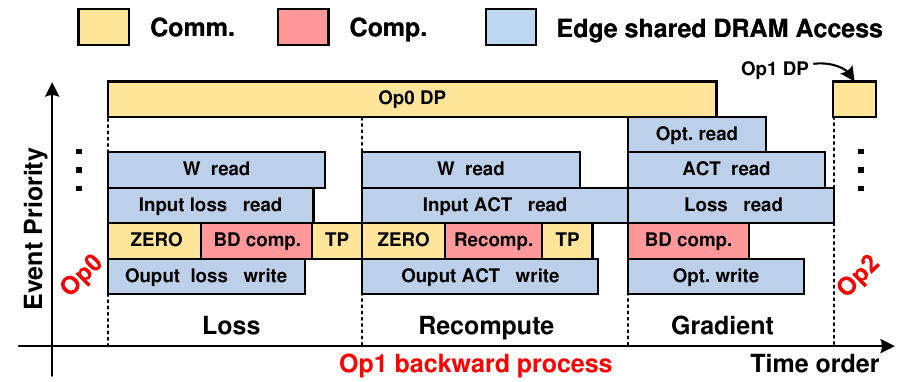}
  \caption{\centering The events in PALM backward process.}
  \label{fig:event}
\end{figure}

In PALM, each operator also generates three types of events: forward, backward, and gradient update. Each type of event is further divided into computation, communication, and memory access tasks. Fig.~\ref{fig:event} describes the main events during backward execution. For each operator, the backward process includes loss computation, activation re-computation, and gradient computation. Activation re-computation occurs only when there is insufficient memory capacity. Each sub-process requires accessing data from memory for computation, with non-negligible communication overhead. The next sub-process begins only after the completion of the current sub-process. For example, in \verb|Recompute| sub-process, we wait for the completion of the \verb|Loss| computation event before entering \verb|Gradient| sub-process. During the three sub-processes, DP communication from the previous operator can overlap with the current operator's execution.
The forward process is similar to the re-computation in the backward process and is not separately listed here.
The main events in the gradient update process only include full-precision weights load from DRAM and store back to DRAM, and we have omitted the accumulation computation in the gradient update process.

\subsection{Adaptive Parallelism Interface}
\ding{182} \textbf{PP}. PP partitions operators of the computation graph into different stages to minimize the pipeline bubble. The \textbf{ideal} execution time in the pipeline training scenario can be evaluated using Eq.~\eqref{pipeline}.
\begin{equation}
\begin{aligned}
ET_{\text{total}} &= (\frac{B}{b}-1)\max_{\text{stages}}(ET_{FD}+ET_{BD}) \\
& + \sum^{\text{stages}}(ET_{FD}+ET_{BD})+ET_{GU},
\end{aligned}
\label{pipeline}
\end{equation}
where $ET$ is executing time, $B$ is the batch size and $b$ is the micro-batch size.
In fact, on the tiled accelerator, the execution time is influenced by the spatial position of the physical tiles corresponding to the stages. We will further discuss this phenomenon with experiments in Section~\ref{position_mapping}. PALM  takes into account that PP results in differences in memory capacity requirements, as discussed in~\cite{CS}. Considering a training pipeline with $S$ stages, activations from each stage are stored in the \texttt{FD} process, until they are consumed for \texttt{GU} in the \texttt{BD} process. For example, the first stage should store $S$ times the activation in 1F1B, and $B$ times the activation in GPipe as illustrated in Fig.~\ref{fig:schedule1}. 
Incorporating the aforementioned considerations into PP modeling, PALM supports users to bind stages based on tile IDs and op IDs with \textbf{\textit{Adaptive Parallelism Interface}} in Fig.~\ref{fig:PALM}, and provides a default way for DL models to allocate stages based on computing power requirements.

\ding{183} \textbf{TP and DP}. 
We analyzed the communication size of all-reduce generated by TP and DP strategies in common operators, as shown in Table~\ref{parallelism}. PALM partitions mapped physical tile groups into communication groups, automatically inserting collective communication events into the tile group event queue. Taking the simple linear operator as an example: The linear operator $Y=WX^T$ has four dimensions $(B, M, N, K)$, where $B$ represents batch size, $K$ represents the reduce dimension, $M,N$ represents the output dimension, $X^{(N\times K)}$represents input, $W^{(M\times K)}$ represents weights, and $Y^{(M\times N)}$ represents output. The dimensions $(b, m, n, k)$ represent the parallelism degree for each corresponding dimension. If we map the operator onto 16 tiles from 0 to 15, it is essential to ensure that $b\times m\times n\times k=16$. The parallelism strategy can be configured by the user as $(2, 2, 2, 2)$ or $(4, 4, 1, 1)$, and so on. Further, corresponding communication groups are automatically generated. During the \verb|FD|, \verb|BD|, and \verb|GU| processes, there is a need for collective communication in the corresponding tile groups. The parallelism of other operators like Conv2 and Pool are the same.
For simplicity, we assume that the input shape of Conv2 or Pool is $(B,C,I,I)$, the shape of weight is $(W,W,K)$, and the shape of output is $(B,K,O,O)$. Specially, $K$ is equal to 1 in Pool operator. The communication size of all-reduce is also represented in Table~\ref{parallelism}. For transformer operator, it is a combination of a series of linear operators. And the shapes of input and output are $(B,S,H)$. We support both DP ($N_d$) and TP ($N_m$) as described by Megatron\cite{shoeybi2019megatron}. The communication size generated by splitting these linear operators is accumulated. These parallelism dimensions such as $(b, m, n, k)$ and $(N_d, N_m)$ can also be configured by the user with the interface in Fig.~\ref{fig:PALM}.

\begin{table*}[b]{\small}\renewcommand{\arraystretch}{1.4}
\caption{Parallelism analysis of common operators.}
\begin{center}\label{parallelism}
\footnotesize
\renewcommand{\arraystretch}{1.5} 
\begin{tabular}{c|c|c|c|c}
\hline
\begin{tabular}[c]{@{}c@{}}\textbf{Operator }\\ \textbf{Type}\end{tabular} & \begin{tabular}[c]{@{}c@{}}\textbf{Dimension}\\ \textbf{Symbol}\end{tabular}& \begin{tabular}[c]{@{}c@{}}\textbf{Parallelism}\\ \textbf{Symbol}\end{tabular} & \begin{tabular}[c]{@{}c@{}}\textbf{Computation Count}\\ \textbf{(FLOPs)}\end{tabular} &\begin{tabular}[c]{@{}c@{}}\textbf{FD;BD;GU }\\ \textbf{(comm. size, comm. dim.)}\end{tabular} \\
\hline
$Linear$ & [B,M,N,K] & (b,m,n,k) & $ \frac{2B  M N  K}{b  m  n  k} $ & $(\frac{BMN}{bmn},k);(\frac{BMK}{bmk},n);(\frac{NK}{nk},b),(\frac{NK}{nk},m)$ \\
\hline
$Conv2$ & [B,H,W,C,R,S,K] & (b,c,i,k) & $\frac{2B  H WRSCK }{bick} $ & $(\frac{BH_OW_OK}{bik},c);(\frac{BHWC}{bic},k);(\frac{RSCK}{ck},b),(\frac{RSCK}{ck},i)$ \\
\hline
$Pool$ & [B,H,W,C,R,S] & (b,c,i,1) & $\frac{2B H W RS C }{bc i }$ & N/A\\
\hline
$Transformer$ & [B,H,S,A] & ($N_d,N_m$,1,1)$^{1}$ &$\frac{24BSH^2 +4BS^2H}{N_d N_m}$ & $(\frac{2BSH}{Nd},Nm);(\frac{2BSH}{Nd},Nm);(\frac{12H^2}{Nm},Nd)$ \\
\hline
\end{tabular}
\end{center}
\begin{tablenotes}
\item[1]$^1$Megatron\cite{shoeybi2019megatron}
\end{tablenotes}
\end{table*}

\begin{algorithm}[htbp]\scriptsize
\caption{\texttt{SRAM allocation method.}}
\label{alg:strategy}
\begin{algorithmic}
\STATE \textbf{Input:} $\mathbf{allocated\_tiles}, \mathbf{split\_ops\_by\_parallelism}$\;  
\STATE $Wt\gets 0,WSG \gets 0,ACT \gets 0$\;  
\FOR{$(i, Op)$ \textbf{in} $split\_ops\_by\_parallelism$}  
    \STATE $Wt \gets Wt + Op.Wt $\;  
    \STATE $WSG \gets WSG + Op.Wt +Op.Optimizer\_State+Op.Gradient$\;  
    \STATE $ACT \gets ACT + Op.I$\;  
\ENDFOR  
\STATE $S\_Cap.\gets SRAM\_Capacity\_Size$\;  
\STATE $Op\_Fd\_Access\_Size \gets [0,...,0]$\;  
\FOR{$(i, Op)$ \textbf{in} $split\_ops\_by\_parallelism$}  
    \IF{$Wt \leq S\_Cap.$}  
        \STATE $strategy \gets \mathbf{activation\_stream}$\;  
        \STATE $Op\_Fd\_Access\_Size[i] \gets Op.I+Op.O$\; 
    \ELSIF{$WSG \leq S\_Cap.$}  
        \STATE $strategy \gets \mathbf{weight\_stream}$\;  
        \STATE $Op\_Fd\_Access\_Size[i] \gets Op.Wt$\;  
    \ELSE  
        \STATE $\Phi_1=\lceil\frac{Op.W}{S\_Cap.}\rceil\times Op.I$\;
        \STATE $\Phi_2=\lceil\frac{Op.I}{S\_Cap.}\rceil\times Op.Wt$\; 
        \IF{$\Phi_1<\Phi_2$}
        \STATE $strategy \gets \mathbf{weight\_stationary}$\;  
        \STATE $Op\_Fd\_Access\_Size[i]=\Phi_1+Op.O$\; 
        \ELSE
        \STATE $strategy \gets \mathbf{input\_stationary}$\; 
        \STATE $Op\_Fd\_Access\_Size[i]=\Phi_2+Op.O$\; 
        \ENDIF 
    \ENDIF  
\ENDFOR  
\STATE\textbf{Output:} $\mathbf{strategy}, \mathbf{Op\_Fd\_Access\_Size}$  
\end{algorithmic}
\end{algorithm}
\subsection{Detailed  Bandwidth Model}
\ding{182} \textbf{SRAM allocation}\label{sgy}.
PALM holds the view that SRAM primarily influences DRAM access. Alg.~\ref{alg:strategy} explains the main modeling idea: Operators are split by the parallelism strategy and their corresponding tiles are taken as input to obtain the corresponding SRAM strategy and  DRAM  access size in forward process. Strategies {$S_{WSG\_ACT}$, $S_{WSG}$, $S_{ACT}$, $S_{PTY}$} respectively represent {weights, optimizer states, weight gradients} (WSG) and input/output activation ($ACT_{IN}, ACT_{OUT}$) either statically stored in on-chip SRAM, one of them stored in on-chip SRAM, or none stored on-chip. 
It is worth noting that when WSG and ACT cannot be retained in SRAM for a long time, PALM adopts a penalty strategy $S_{PTY}$, modeling extra DRAM accesses for WSG and ACT.  When $ACT \geq W$, we use input stationary (IS), otherwise, we use weight stationary (WS).
PALM considers storage differences brought about by the optimizer. For optimizer Adam, it requires storage for first-order and second-order moments related to weights, and gradients of backward activations, significantly increasing storage requirements. If optimizer SGD is used, there is no overhead for optimizer states. During inference, there is no storage overhead for gradients.
Alg.~\ref{alg:strategy} only lists the DRAM access size for the forward process. The analysis for the backward and gradient update process follows the same methodology, thus being neglected here.

\ding{183} \textbf{Detailed NoC model}.
The ideal communication latency of the NoC can be obtained using Eq.~\eqref{noc}, where  $Link\_Time$ represents single hop link delay and $ Hops$ represents the total number of hops in the communication path. However, the analytical model~\cite{won2023astra} does not consider whether \textbf{all} links are idle at a given moment in a transmission path. Hence, the specific latency of \textbf{contention\_delay} can not be obtained by the analytical method. In the presence of congestion, the communication time may degrade to Eq.~\eqref{noc_congestion} in the analytical model, which means a hop-by-hop data transmission, without forming a pipeline transmission along the link. But it is equivalent to reducing the bandwidth of the NoC by $Hops$ times.
Even the modeling of the latter cannot guarantee that the single hop transmission is not occupied by other tasks.
\begin{equation}
\begin{aligned}
Comm\_Time& =Link\_Time\times Hops +\frac{Comm\_Size}{BW_{Link}} \\
& +\mathbf{Contention\_delay},
\end{aligned}
\label{noc}
\end{equation}
\begin{equation}
\begin{aligned}
Comm\_Time& =(Link\_Time+\frac{Comm\_Size}{BW_{Link}})\times Hops.
\end{aligned}
\label{noc_congestion}
\end{equation}

PALM considers NoC congestion, treating the link as an exclusive resource during execution. When a link is occupied by the current task, the execution time can be obtained by Eq.~\eqref{noc}. Communication tasks can only be executed when needed link are not occupied. Otherwise, they will wait for the release of resources.

\ding{184} \textbf{Detailed DRAM model}. Through the analysis of SRAM, the size of DRAM memory access has been determined, and ideally, the memory access latency $Access\_Time$ can be obtained using Eq.~\eqref{eq:access}. 
However, in the tiled accelerator, the DRAM is shared among tiles. Due to the varying distances of different tiles from DRAM and the different times they initiate memory access requests, the understanding of whether the bandwidth ($BW_{DRAM}$) is occupied at a particular moment is not clear enough. Eq.~\eqref{eq:access} cannot accurately represent memory access latency.
\begin{equation}
\begin{aligned}
Access\_Time=Response\_Time+\frac{Access\_Size}{\mathbf{BW_{DRAM}}}.
\end{aligned}
\label{eq:access}
\end{equation}
\begin{equation}
\begin{aligned}
&DRAM\_Time=Access\_Time+NoC\_Time.
\end{aligned}
\label{eq:dram}
\end{equation}

Based on the above equation, PALM constructs a memory access model for edge-shared DRAM in tiled accelerators. PALM considers DRAM bandwidth as a resource that is occupied during execution like the NoC model. The data transmission time, denoted as $NoC\_Time$, through the NoC has been taken into account. Therefore, the total DRAM access time $DRAM\_Time$ of a tile can be obtained using Eq.~\eqref{eq:dram}.


\section{Case Study}\label{experiment}
\subsection{Verification of Simulation Accuracy }
\subsubsection{\textbf{Verification of NoC model and DRAM model}}
\begin{figure}[t]
  \centering
  \subfloat[\centering 4 devices in ring topology]{
    \includegraphics[width=0.49\linewidth]
    {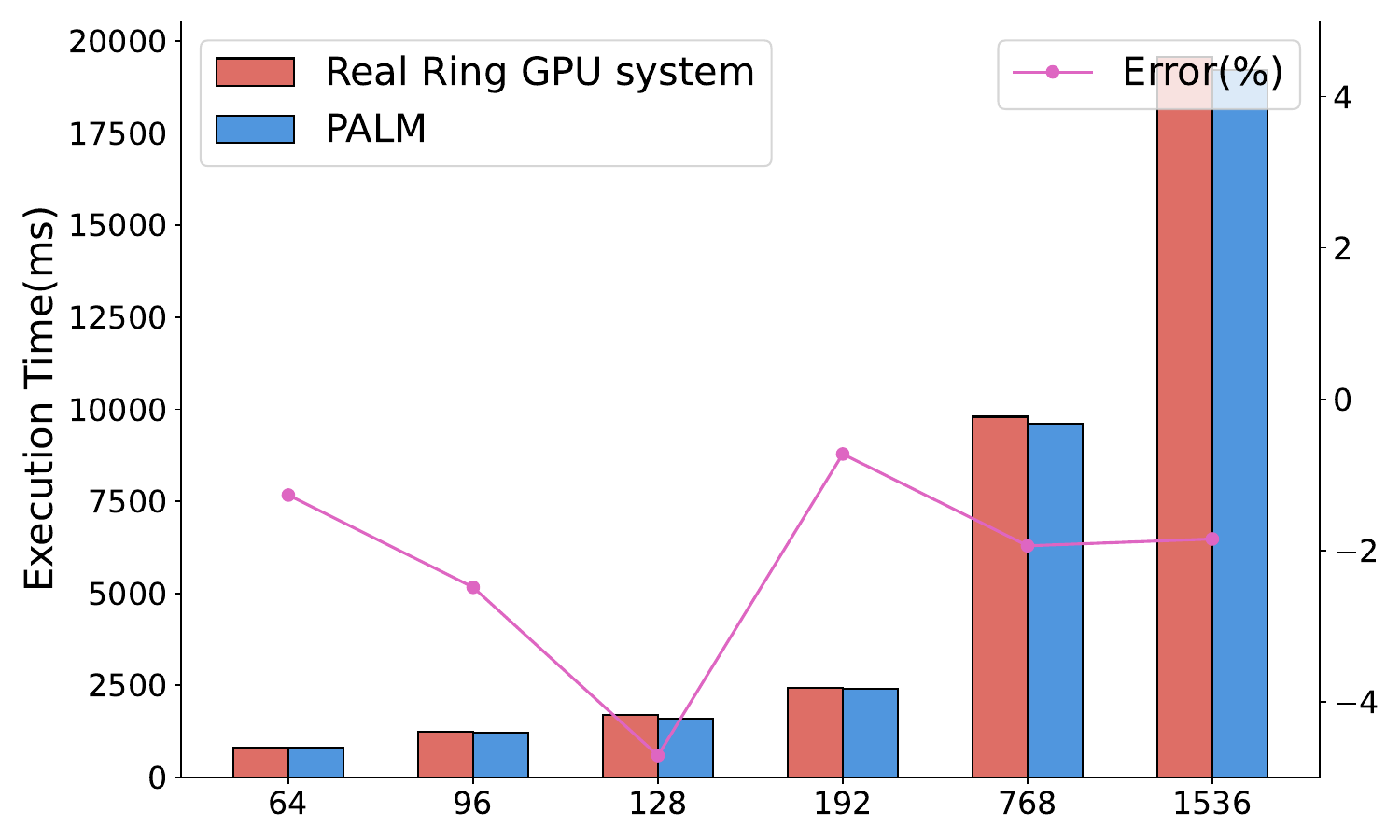}\label{fig:test_gpu_ring_4}}
  \subfloat[\centering   16 devices in ring topology]{
    \includegraphics[width=0.49\linewidth]
    {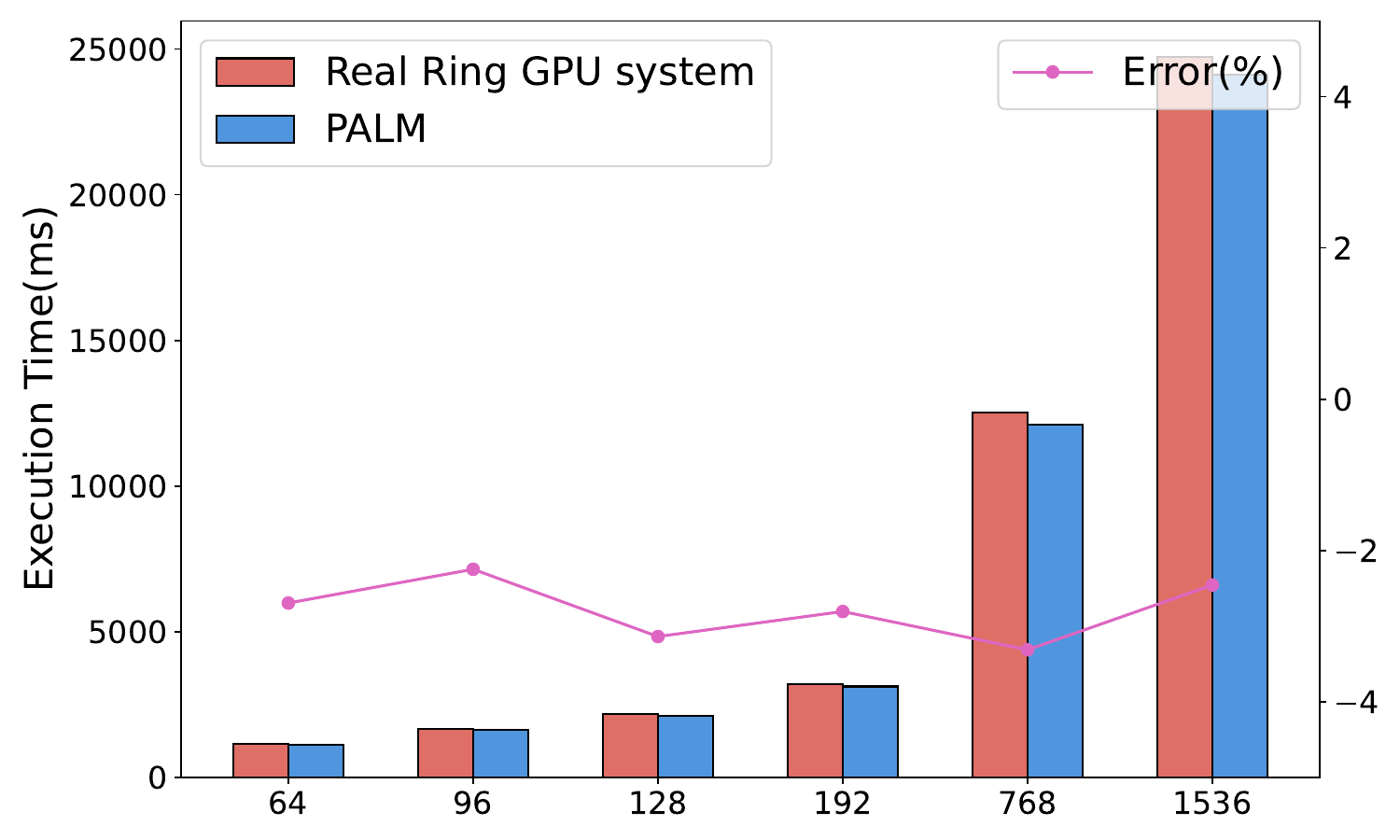}
    \label{fig:test_gpu_ring_16}
  }
  \caption{ Performance comparison of PALM simulator with GPU system for the all-reduce task under ring topology .}
  \label{fig:all-reduce}
\end{figure}

\begin{figure}[t]
    \centering
    
        \includegraphics[width=0.8\linewidth, height=0.5\linewidth]{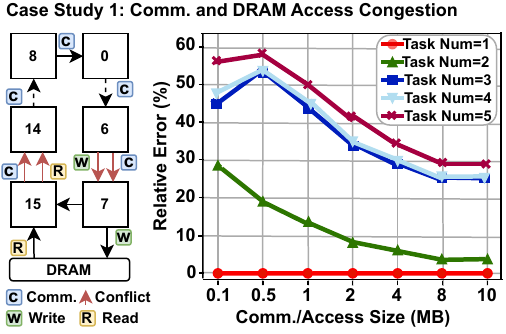}
        \caption{Error of multi-task stacked on tiled accelerator in PALM VS Analytical model.}
        \label{fig:congestion}
    
\end{figure}


    

To validate NoC model, we conduct the base ring all-reduce task on PALM. As depicted in Fig.~\ref{fig:all-reduce}, the error on 4 and 16 tiles is within 5\%, compared with the results from a real GPU system with ring topology in \cite{won2023astra}. 

To validate the congestion phenomenon, we conduct experiments in Fig.~\ref{fig:congestion} involving all-reduce, all-to-all, and DRAM read and write tasks overlapping, where we use a different number of task combinations. 
The results show that the execution time of the analytical model is at most 50\% less than that of the congestion model. When the number of tasks is 5 and the single task communication/access size is 8MB, the execution time of the analytical model is 30\% less, and it stabilizes at this value as the communication/access  increases. According to the previous analysis, these numerical differences reflect the modeling error of the analytical model. Therefore, it can be proven that PALM modeling tasks are necessary for congestion scenarios.

\subsubsection{\textbf{Verification of Scheduling and Parallelism }}
Because of the limited LLM data for tiled architecture, we collect published LLM data from GPU cluster to validate the scheduling and parallelism analysis. We replace the underlying 2D topology of PALM with GPU topology. The result in Table~\ref{gpu_sim} indicates that the average total error of PALM scheduling and parallelism analysis is less than 15\%.
\subsubsection{\textbf{Verification on tiled accelerator}}
We use PALM to simulate the ResNet50 and Bert-base inference task on Tenstorrent Grayskull~\cite{vasiljevic2021compute} architecture. By adjusting the mapping strategy, our simulated throughput has an error of less than 13\% compared to the published throughput as shown in Table~\ref{tenstorrent}. In pipeline inference, there is continuous data input without a backward process. Therefore, we obtain throughput that ignores the pipeline drain time and setup time as illustrated in Fig.~\ref{fig:schedule1}.

\begin{table}[t]\renewcommand{\arraystretch}{1.0}\caption{ Performance comparison of \\ PALM and Megatron published data.}\label{gpu_sim}

\begin{tabular}{c|c|c|c|c}
\toprule
\textbf{Model} & \textbf{TP, DP, PP} & \textbf{PALM seq/s} & \textbf{Published seq/s$^{1}$} & \textbf{Error \%} \\
\midrule
T-18B & 8, 32, 1 & 114.294 & 116.415 & 1.82 \\
T-39B & 8, 32, 2 & 100.230 & 111.565 & 10.16 \\
T-76B & 8, 32, 4 & 96.601 & 115.898 & 15.65 \\
T-145B & 8, 24, 8 & 83.888 & 95.720 & 12.36 \\
T-310B & 8, 15, 16 & 51.140 & 58.738 & 12.94 \\
T-530B & 8, 9, 35 & 40.007 & 47.440 & 15.60 \\
\bottomrule
\end{tabular}
    \begin{tablenotes}
    \item $^1$ Performance with mixed precision training.
    \end{tablenotes}
    
\end{table}

\begin{table}[t]\renewcommand{\arraystretch}{1.15}
\begin{threeparttable}
\centering
\caption{Performance comparison of \\
PALM and Grayskull published data. }\label{tenstorrent}

\begin{tabular}{c|c|c|c}
\specialrule{0.12em}{0.1pt}{0.2pt}
\textbf{Model name} &\textbf{PALM sample/s}& \textbf{Published sample/s}& \textbf{Error \%} \\
\hline
ResNet50 & 23033.46 & 22431$^{1}$ ~\cite{gwennap2020tenstorrent} & 2.68 \\
Bert-base &  3190.12 & 2830 ~\cite{vasiljevic2021compute} & 12.72\\
\specialrule{0.12em}{0.1pt}{0.2pt}
\end{tabular}

\begin{tablenotes}
      \item $^1$Performance with int\,8 computing power. 
\end{tablenotes}
\end{threeparttable}
\end{table}

\subsection{ Parallelism of LLM on Wafer-scale Architecture}
We explore the influence of wafer-scale architecture on the optimal parallelism of LLM.
Based on PALM, we build a wafer-scale architecture with specific parameters, as shown in Table~\ref{tab:wafer_scale}. The overall system consists of a $5\times4$ tile array with $4\times4$ core per tile, communicated with tile-to-tile and core-to-core NoC. 
We have selected models T-18B, T-76B, and T-145B as the baseline in Table~\ref{gap}, with (TP=8, DP=2, PP=20). The performance of the baseline is close to the result presented in Table~\ref{gpu_sim}.
\subsubsection{\textbf{Optimal parallelism analysis}}
\begin{figure}[t]\renewcommand{\arraystretch}{1.15}
  \centering
  \includegraphics[width=1\linewidth]{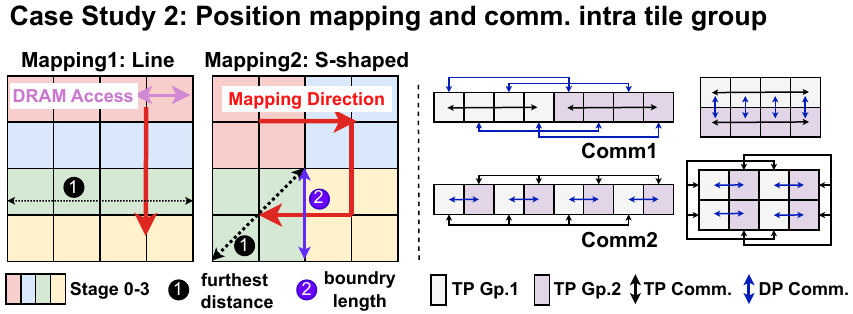}
  \caption{Position mapping in inter-tile groups and communication strategies in intra-tile groups.}
  \label{fig:layout}
\end{figure}

\begin{figure}[t]
  \centering
  \subfloat[\centering mapping1$+$comm2 strategy]{
    \includegraphics[width=0.48\linewidth, height=0.3\linewidth]{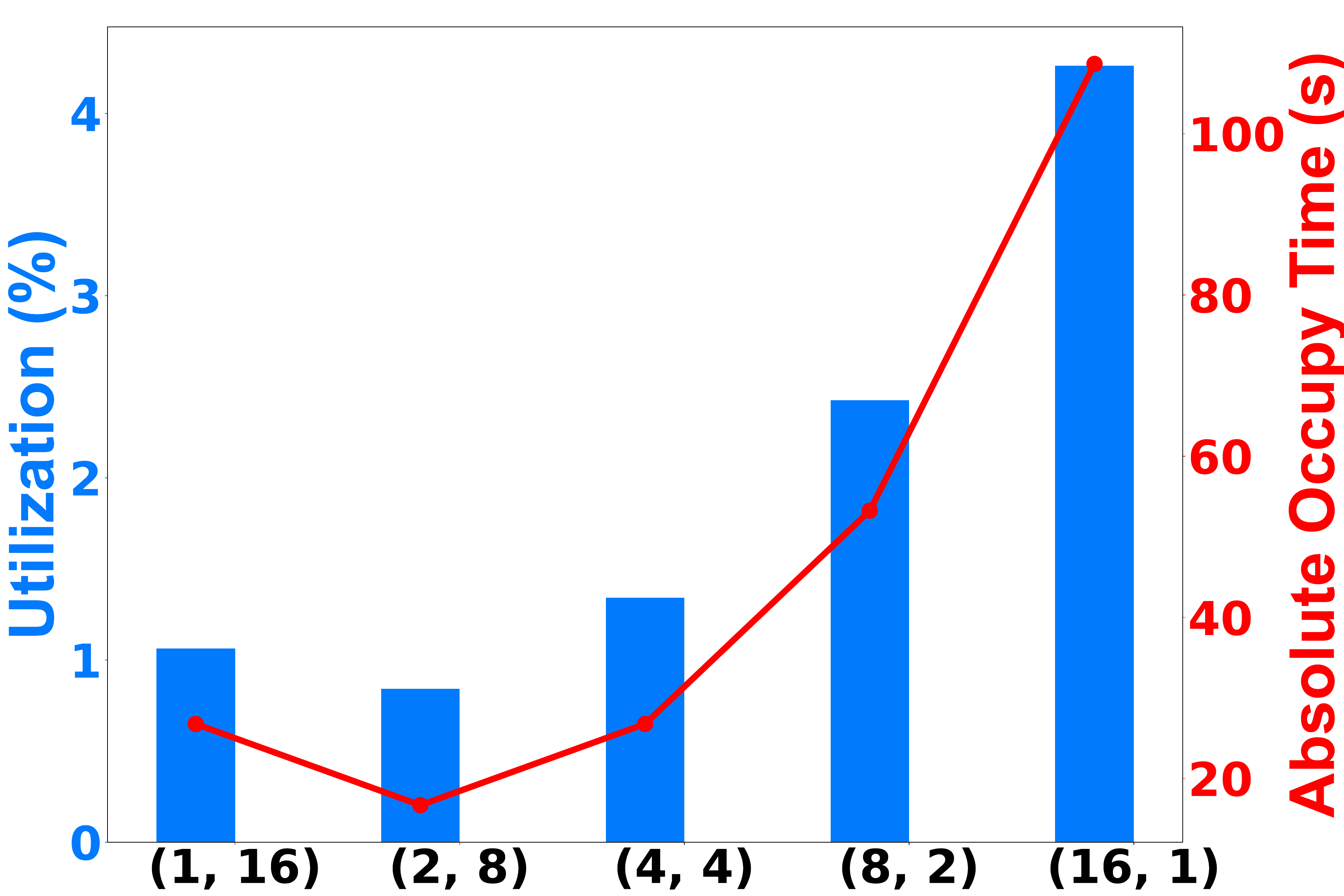}\label{fig:time_util}}
  \subfloat[\centering mapping2$+$comm1 strategy]{
    \includegraphics[width=0.48\linewidth, height=0.3\linewidth]{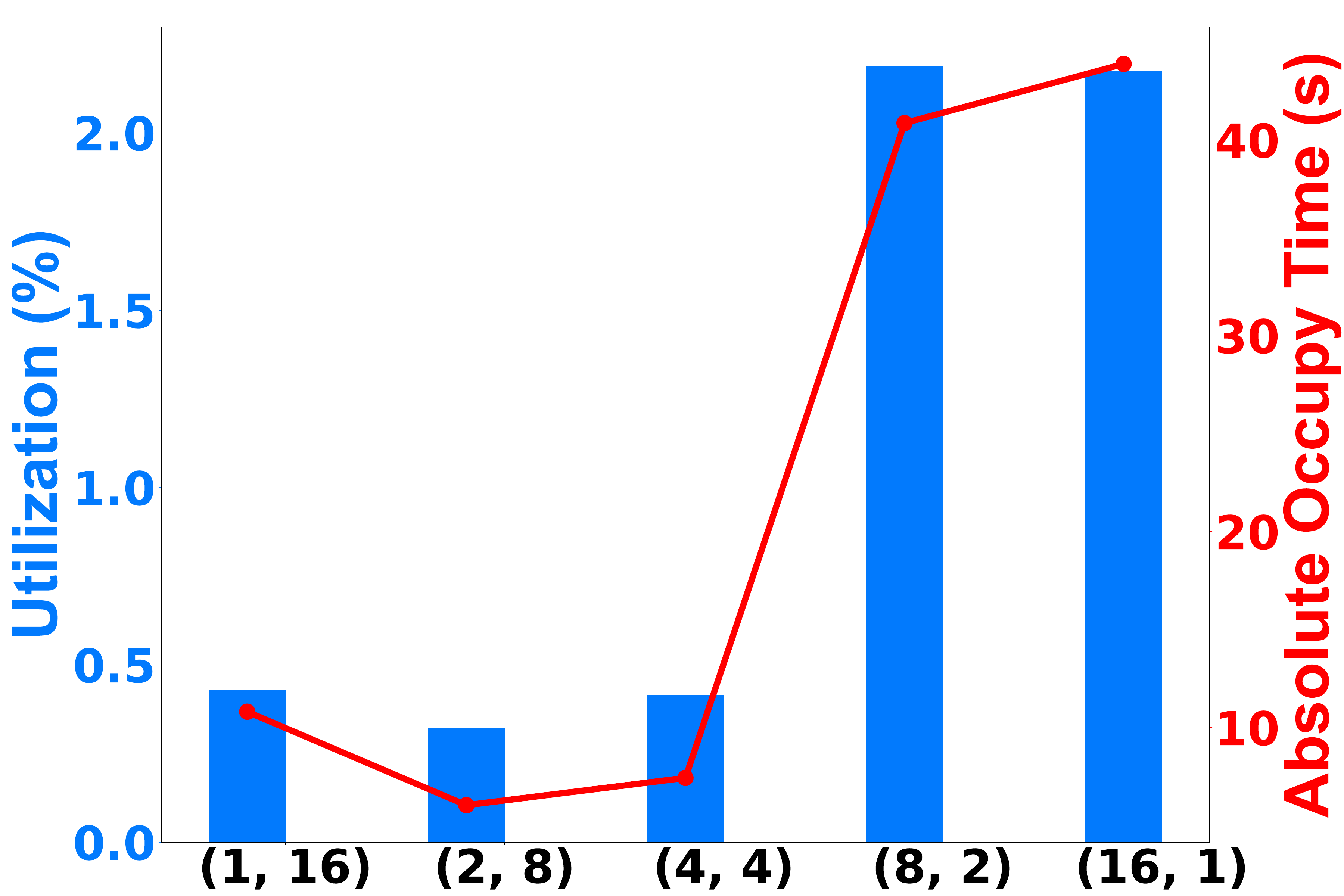}
    \label{fig:time_util1}
  }
  \caption{The average utilization and absolute occupy time of NoC on wafer-scale architecture for T-145B task.}
  \label{fig:util}
\end{figure}
For a single transformer operator, the total communication size is determined by Eq.~\eqref{transformer}, which influences the communication latency at the top level. 
\begin{equation}
\begin{aligned}
Comm\_Size=\frac{8BSHN_m}{N}+\frac{24H^2}{N_m},
\end{aligned}
\label{transformer}
\end{equation}

\noindent where $B$, $S$, and $H$ represent the model parameters. $N_m$ represents the degree of TP, and $N$ represents the degree of DP multiplied by TP. In this experiment, $N$ is set to 16. To minimize communication size, the optimal value for $N_m$ is 1.6, close to 2. The optimal throughputs shown in Fig.~\ref{fig:wafer1} and Fig.~\ref{fig:wafer2} validate this conclusion.

As illustrated in Fig.~\ref{fig:util}, the minimum average NoC occupancy time on T-145B task is consistent with (TP=2, DP=8) to minimize communication size. However, the optimal throughput corresponds to (TP=4, DP=4) as shown in  Fig.~\ref{fig:wafer3}. This indicates that minimal communication size does not always lead to absolute performance optimization, and actual architecture needs to be considered as well.

\subsubsection{\textbf{Impact of position mapping for stage}}\label{position_mapping}
Two common mapping layouts are illustrated in Fig.~\ref{fig:layout}. The line layout arranges the pipeline vertically, with data passing vertically across stages, and intra-stage communication and memory access occurring horizontally. The S-shaped layout considers the trade-off of the furthest distance between mapped tiles and the boundary length of the tile group. In our experiments, the number of layers in the baseline model is the same as the number of tiles, with the $4\times4$ cores in a tile forming one stage. The high bandwidth within the tile supports DP and TP effectively, while inter-tile bandwidth is lower, aligning with the low communication requirements of PP.

\begin{figure*}[tb]
  \subfloat[\centering T-18B]{
  \centering
    \includegraphics[width=0.32\linewidth, height=3.5cm]{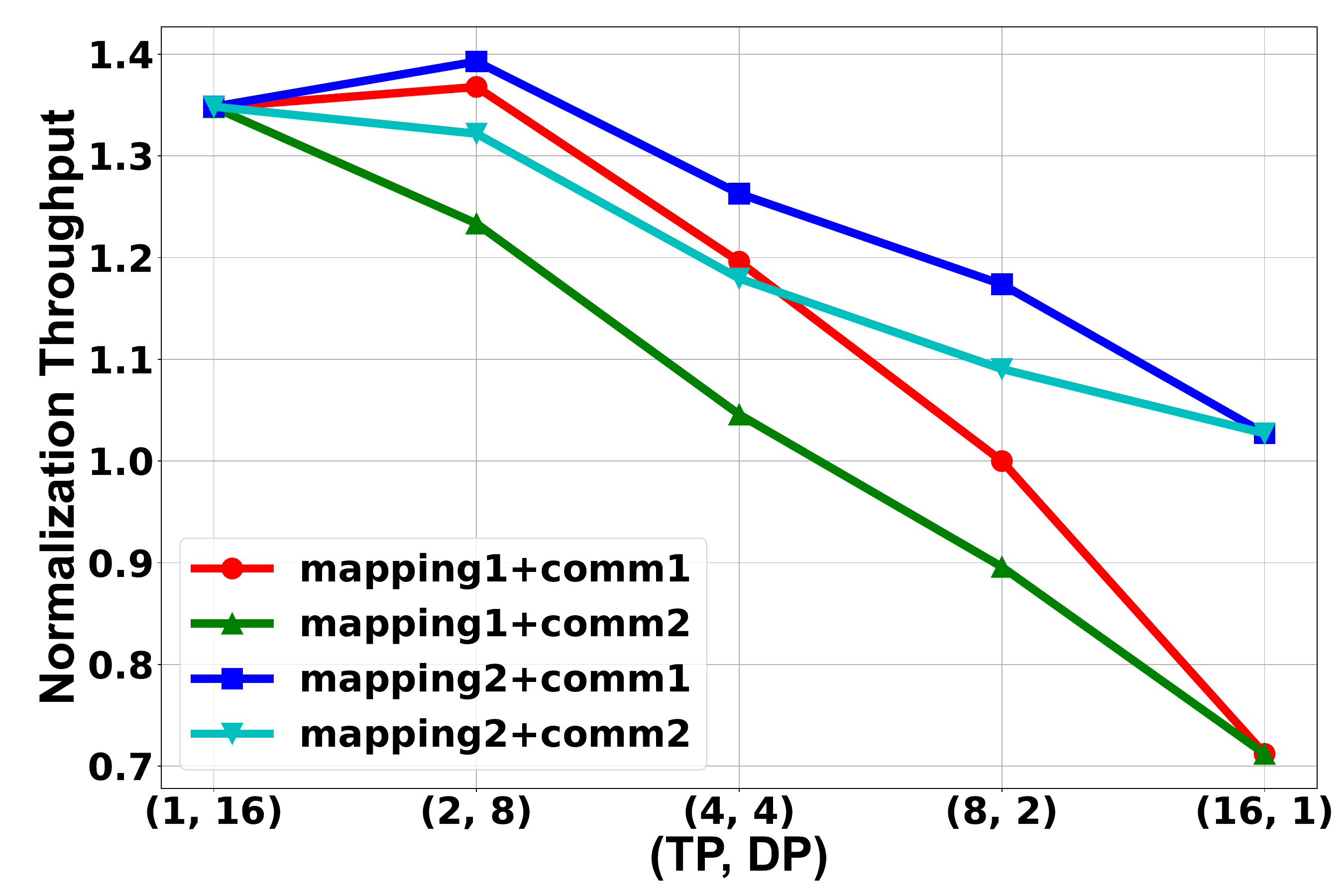}
    \label{fig:wafer1}
  }
  \subfloat[\centering T-67B]{
    \includegraphics[width=0.32\linewidth, height=3.5cm]{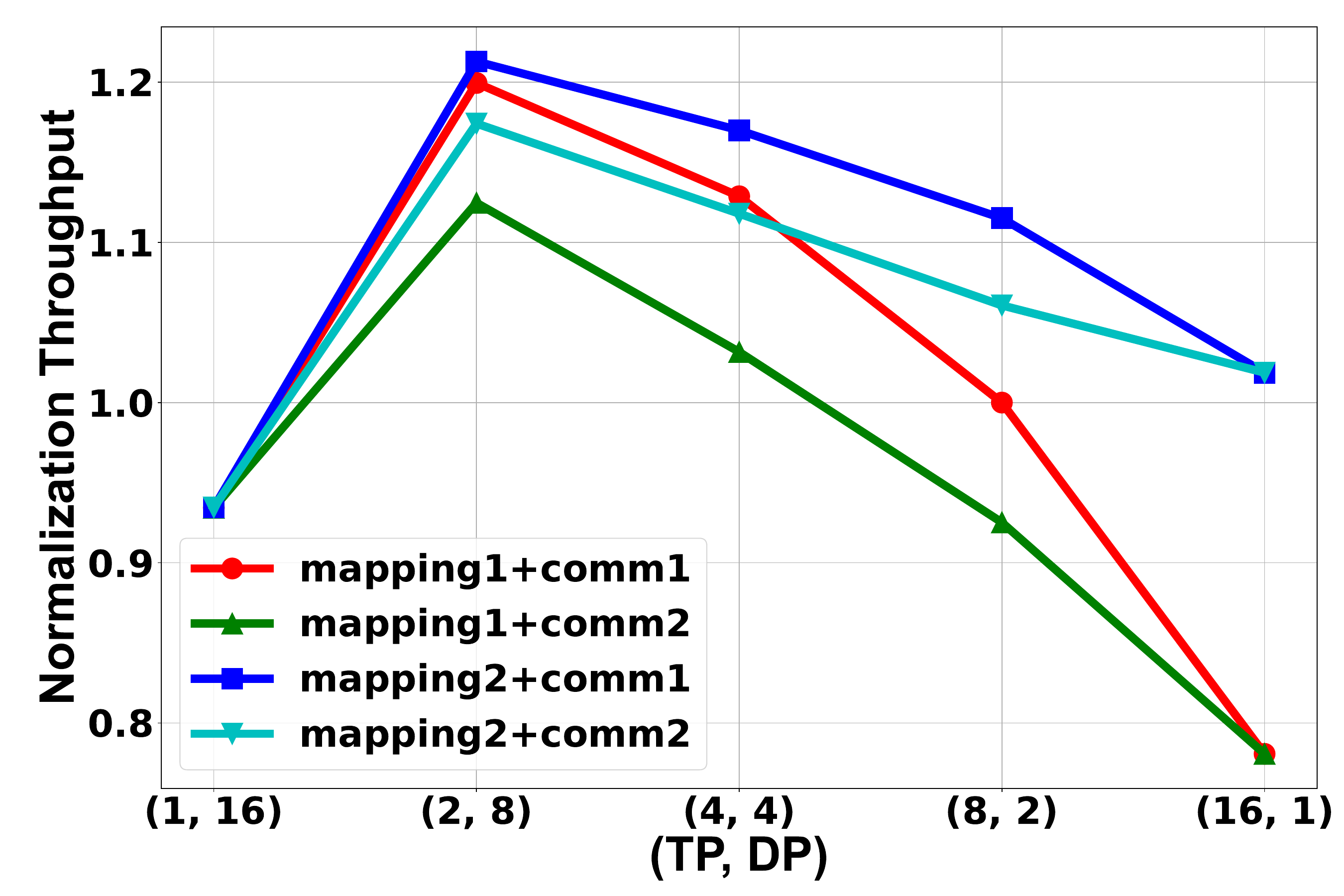}
    \label{fig:wafer2}
  }
  \subfloat[\centering T-145B]{
    \includegraphics[width=0.32\linewidth, height=3.5cm]{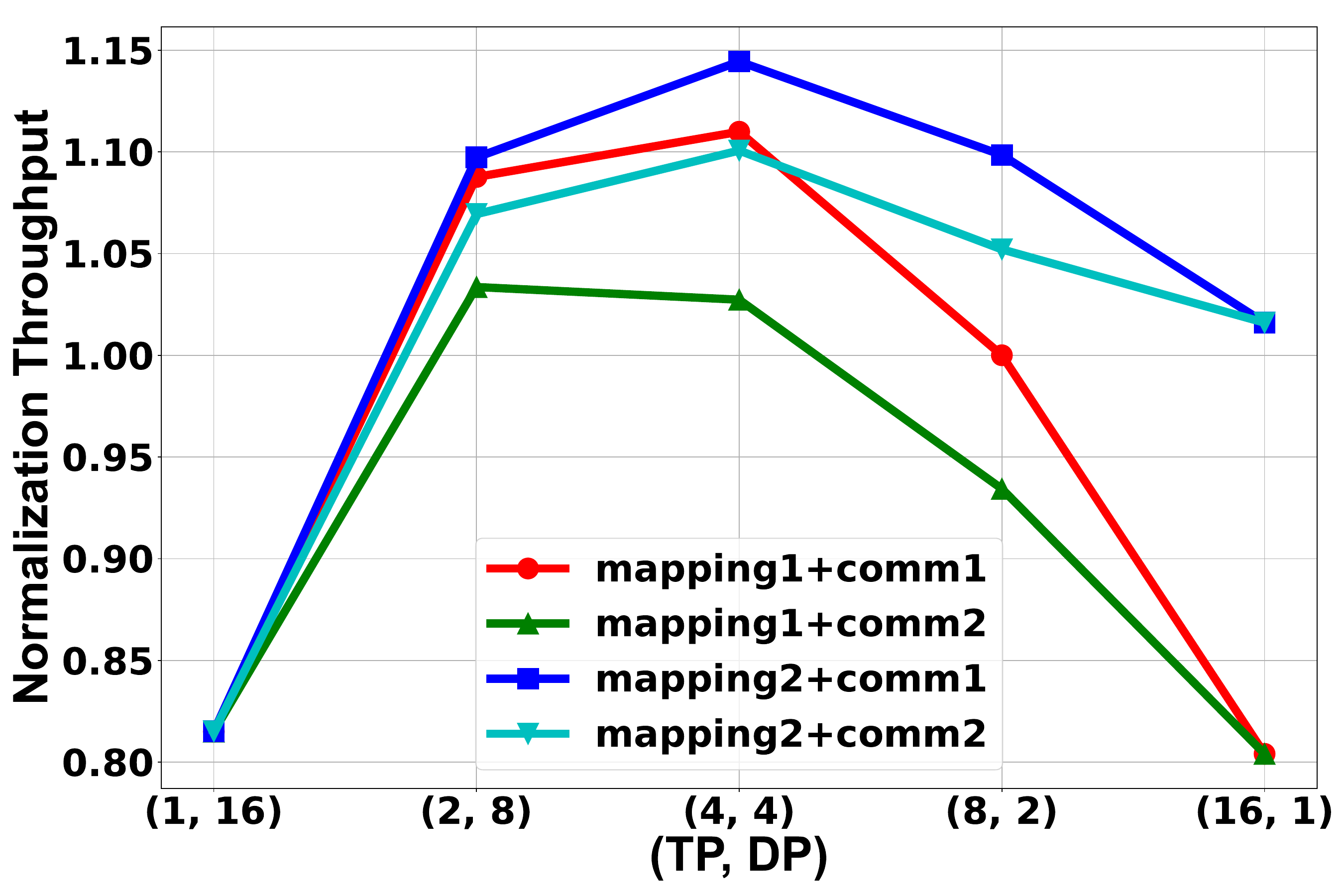}
    \label{fig:wafer3}
  }
  \caption{Performance comparison among various combinations of mapping methods and TP communication strategies on wafer-scale architecture.}
  \label{fig:wafer}
\end{figure*}

Fig.~\ref{fig:wafer} illustrates experimental results, where \verb|mapping1| represents the Line layout, and \verb|mapping2| represents the S-shaped layout. The results validate that the S-shaped layout exhibits better performance.


\subsubsection{\textbf{Impact of communication group in stage}}
\verb|comm1| represents TP communication group as close as possible in topology, \verb|comm2| represents the opposite, which is shown in Fig.~\ref{fig:layout}. Fig.~\ref{fig:wafer} also shows that the performance with \verb|comm1| is better. As analyzed earlier, when TP$\geq$2, the first term in Eq.~\eqref{transformer} contributes to an increasing communication size. Considering the allocation of TP within intra-groups, it is crucial to prioritize minimizing the distance between cores along the TP communication dimension to reduce communication time.

Based on the results, we conclude that the minor optimizing parallelism strategies can lead to at least $2\times$ performance gap. This improvement comprises a 40\% contribution from stage position layout and a 60\% contribution from operator-level parallelism and communication optimization.

\begin{table}[t]
\renewcommand{\arraystretch}{1.0}
\centering
\caption{\centering Wafer-scale Configuration Parameters.}\label{tab:wafer_scale}
\begin{tabular}{l||m{3cm}<{\centering}}
\specialrule{0.12em}{0.1pt}{0.2pt}
\textbf{Computing power of single tile}&256 TFlops@FP16 \\
\hline
\textbf{Capacity of single tile SRAM}  & 60 MB  \\
\hline
\textbf{Number of intra-tiles} & $4\times 4$  \\
\hline
\textbf{Edge shared DRAM per tile} & 256 GB/s  \\
\hline
\textbf{Number of tiles} & $5\times 4$  \\
\hline
\textbf{NoC bandwidth of intra-tile}  & 1024 GB/s  \\
\hline
\textbf{NoC bandwidth of inter-tile} & 256 GB/s  \\
\hline
\textbf{Topology} & 2D-mesh  \\
\specialrule{0.12em}{0.1pt}{0.2pt}
\end{tabular}
\end{table}

\begin{table}[t]\renewcommand{\arraystretch}{1.15}
\begin{threeparttable}
\centering
\caption{ Performance comparison of \\
PALM on wafer-scale with GPU published data.}\label{gap}

\begin{tabular}{c|c|c|c}
\specialrule{0.12em}{0.1pt}{0.2pt}
\textbf{Model name} &\textbf{PALM sample/s}& \textbf{Published sample/s$^{1}$}& \textbf{Gap \%} \\
\hline
T-18B & 7.3457 &7.2760 & 0.9  \\
T-76B & 2.0652 &1.7968& 14.94\\
T-145B & 1.1238  &0.9896 & 13.56\\
\specialrule{0.12em}{0.1pt}{0.2pt}
\end{tabular}

\begin{tablenotes}
      \item $^1$ Linear equivalence based on computational power.  
\end{tablenotes}
\end{threeparttable}
\end{table}

\subsection{Communication Optimization}

Due to the bandwidth limitations of the GPU cluster architecture, there is only a single choice for its communication strategy~\cite{Zhuang2022OnOT}. In wafer-scale systems, close intra- and inter-bandwidth can support different communication strategies to minimize costs. Adapter tiles\cite{james2020ispd} are the tiles within the destination group receiving data from the source tile group.

\begin{figure}[tb]
  \centering
  \includegraphics[width=0.9\linewidth]{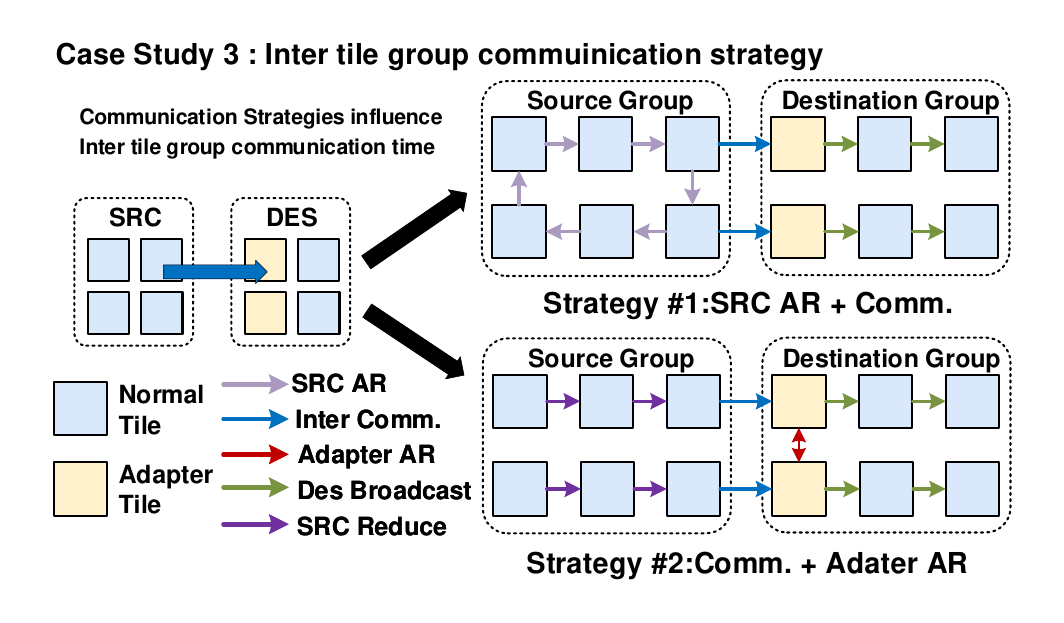}
  \caption{\centering Communication strategies in inter-tile groups. }
  \label{fig:strategy}
\end{figure}

Two communication strategies for inter-tile groups are depicted in Fig~\ref{fig:strategy}. The first involves all-reduce within the source group, data transmission to the destination, and broadcast within the destination. The second reduces the source based on adapters, performs inter-tile transmission, and conducts all-reduce and broadcast in the destination.

Strategy 1's inter-tile communication time is shown by Formula \ref{strategy1}, while Strategy 2's is shown by Formula \ref{strategy2}. In the formulas SG represents the source tile group, DG represents the destination tile group, AR represents all-reduce, R represents reduce, and B represents broadcast. 
\begin{equation}
\begin{aligned}
T=T_{SG\_AR}+T_{Inter\_Comm}+T_{DG\_B}.
\end{aligned}
\label{strategy1}
\end{equation}
\begin{equation}
\begin{aligned}
T=T_{SG\_R}+T_{Inter\_Comm}+T_{Adapters\_AR} + T_{DG\_B}.
\end{aligned}
\label{strategy2}
\end{equation}

Based on BERT-base model, we assess the performance of two communication strategies. The
first set of experiments compares 12 tile source and destination groups under ring shape all-reduce, while the second adds a tile to disrupt ring formation and reassessing performance.


\begin{figure}[tb]
 \centering
  \subfloat[\centering Ring shape]{
    \includegraphics[width=0.49\linewidth]
    {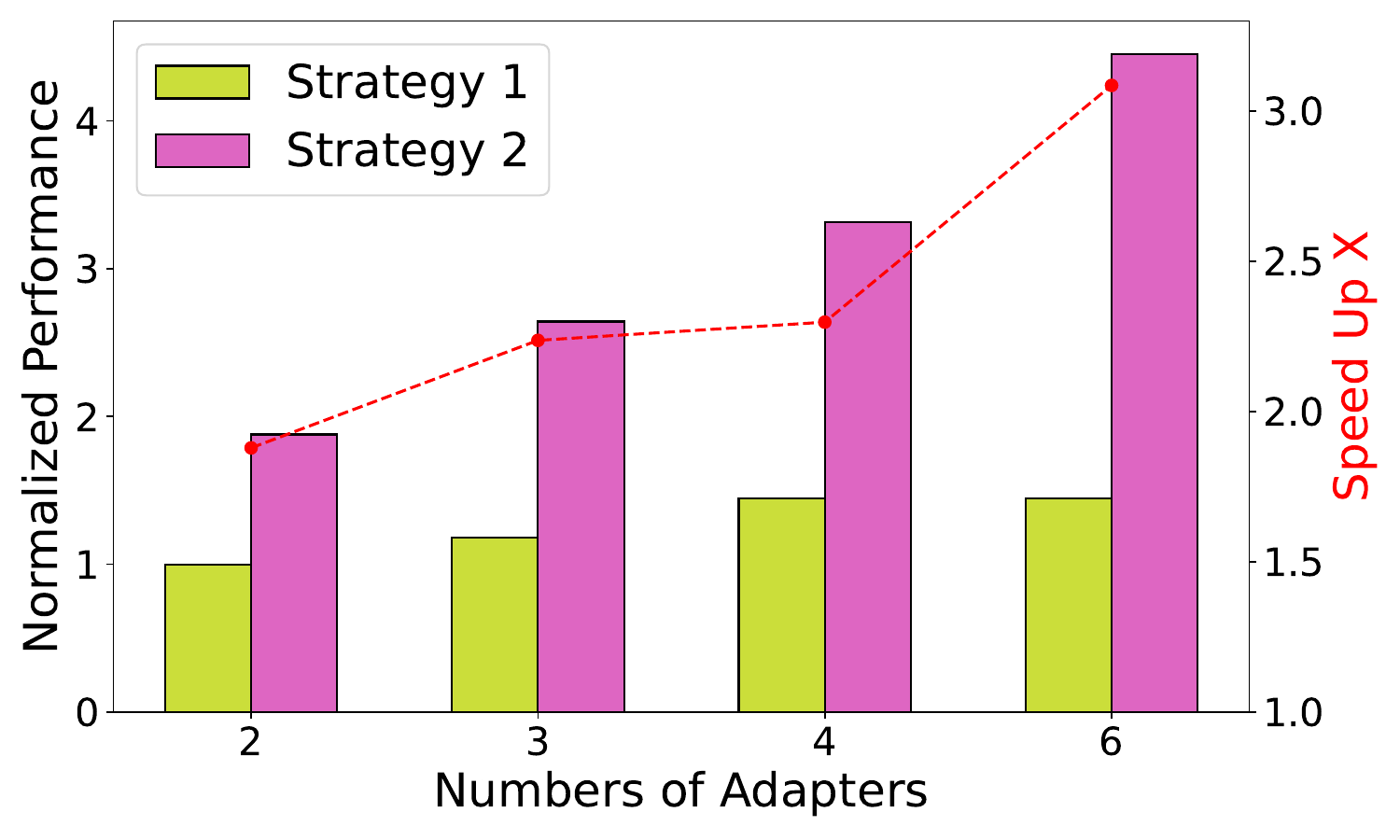}\label{fig:ring-strategy}}
  \subfloat[\centering  Non-ring shape]{
    \includegraphics[width=0.49\linewidth]
    {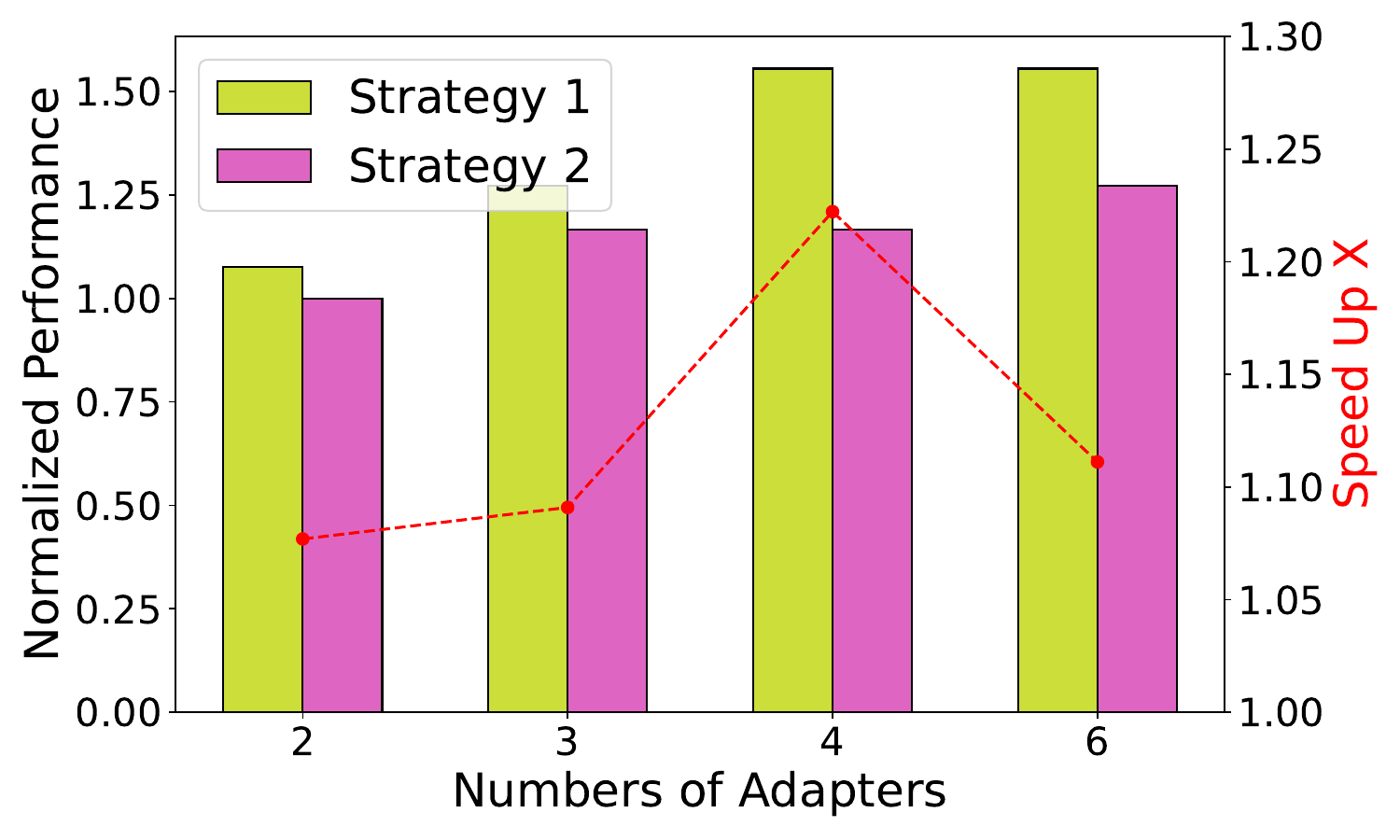}
    \label{fig:no-ring-strategy}
  }
  \caption{Performance of diverse communication strategies in inter-tile groups.}
  \label{fig:casestudy3}
\end{figure}

In Fig~\ref{fig:ring-strategy}, when a ring structure is formed in the source tile group, strategy 1 outperforms strategy 2 in inter-communication performance. This is due to the smaller overall latency of ring all-reduce, resulting in a smaller communication time compared to strategy 2. Moreover, with more adapters participating in inter-communication, the performance of strategy 1 gradually improves by reducing broadcast time in the destination tile group. In Fig~\ref{fig:no-ring-strategy}, when a ring structure cannot be formed, strategy 2 shows better communication performance. In this case, the total time of the reduce and the all-reduce in strategy 2 is smaller than the all-reduce time in the source group of strategy 1. Additionally, the performance of strategy 2 initially improves and then declines as the number of adapters increases, due to the trade-off between the reduce cost and the all-reduce time among adapters. 

According to the result, it is evident that inter-tile communication in ring shape configurations exhibits superior performance under strategy 1, leading to 3.08$\times$ performance gap over strategy 2. Conversely, non-ring shapes are more suitable for the adoption of strategy 2, with a performance increase of approximately 1.23$\times$ compared with strategy 1.

\section{Related Work}\label{sec:Related_work}
There have been multiple arts aimed at predicting the performance of training workload in deep learning. Works~\cite{kosson2021pipelined,liu2022autopipe} were devoted to designing an automatic planner to partition the workload more evenly, aiming at reducing the pipeline bubble time. Moreover, Diksha \emph{et al.} provided an analytical model to predict the training time targeting distributed Transformer~\cite{moolchandani2023amped}. Rasshidi~\emph{et al.} proposed a simulator named Astra-Sim~\cite{rashidi2020astra}, for hardware-software co-design exploration of deep learning training. However, the Astra-Sim mainly focused on examining the impact of varied network topologies and neglects the support for arbitrary parallelism. To this end, its improved version Astra-Sim~2.0~\cite{won2023astra} was proposed to further provide a mechanism to represent and study arbitrary multi-dimensional topologies at scale, with different shapes and bandwidth configurations. However, all the works mentioned above fail to model the space property for tiled accelerators. Though work ~\cite{cai2023inter} designed an inter-layer scheduling space and exploration framework for tiled accelerators, it focused on DNN inference and operator mapping, instead of performance evaluation for DNN training. 

\section{Conclusion}
We propose PALM, a simulator for evaluating tiled accelerators and even wafer-scale architecture in DL training. We consider multiple dimensions that impact training, such as pipeline scheduling, parallelism, tile dataflow, NoC congestion, and so on. Using PALM, we evaluate the training and inference performance throughput of LLM and ResNet models under several tiled accelerators. Compared with the published data, our result has an error of less than 16\%. We discuss the spatial optimization problem of parallelism strategy and communication.
We hope that this work will be further refined in the future to guide subsequent research on mapping algorithms and tiled accelerator design.

\bibliography{ref/refs}

\begin{thebibliography}{10}

\bibitem{he2016deep}
Kaiming He, Xiangyu Zhang, Shaoqing Ren, and Jian Sun.
\newblock Deep residual learning for image recognition.
\newblock In {\em Proceedings of the IEEE conference on computer vision and
  pattern recognition}, pages 770--778, 2016.

\bibitem{carion2020end}
Nicolas Carion, Francisco Massa, Gabriel Synnaeve, Nicolas Usunier, Alexander
  Kirillov, and Sergey Zagoruyko.
\newblock {End-to-End Object Detection with Transformers}.
\newblock In {\em European conference on computer vision}, pages 213--229.
  Springer, 2020.

\bibitem{dosovitskiy2020image}
Alexey Dosovitskiy, Lucas Beyer, Alexander Kolesnikov, Dirk Weissenborn,
  Xiaohua Zhai, Thomas Unterthiner, Mostafa Dehghani, Matthias Minderer, Georg
  Heigold, Sylvain Gelly, Jakob Uszkoreit, and Neil Houlsby.
\newblock {An Image is Worth 16x16 Words: Transformers for Image Recognition at
  Scale}.
\newblock {\em arXiv preprint arXiv:2010.11929}, 2020.

\bibitem{liu2022swin}
Ze~Liu, Han Hu, Yutong Lin, Zhuliang Yao, Zhenda Xie, Yixuan Wei, Jia Ning, Yue
  Cao, Zheng Zhang, Li~Dong, Furu Wei, and Baining Guo.
\newblock {Swin Transformer V2: Scaling up Capacity and Resolution}.
\newblock In {\em Proceedings of the IEEE/CVF conference on computer vision and
  pattern recognition}, pages 12009--12019, 2022.

\bibitem{liu2021swin}
Ze~Liu, Yutong Lin, Yue Cao, Han Hu, Yixuan Wei, Zheng Zhang, Stephen Lin, and
  Baining Guo.
\newblock {Swin Transformer: Hierarchical Vision Transformer using Shifted
  Windows}.
\newblock In {\em Proceedings of the IEEE/CVF international conference on
  computer vision}, pages 10012--10022, 2021.

\bibitem{rombach2022high}
Robin Rombach, Andreas Blattmann, Dominik Lorenz, Patrick Esser, and Bj{\"o}rn
  Ommer.
\newblock {High-Resolution Image Synthesis with Latent Diffusion Models}.
\newblock In {\em Proceedings of the IEEE/CVF conference on computer vision and
  pattern recognition}, pages 10684--10695, 2022.

\bibitem{radford2018improving}
Alec Radford, Karthik Narasimhan, Tim Salimans, and Ilya Sutskever.
\newblock {Improving Language Understanding by Generative Pre-Training}.
\newblock 2018.

\bibitem{raffel2020exploring}
Colin Raffel, Noam Shazeer, Adam Roberts, Katherine Lee, Sharan Narang, Michael
  Matena, Yanqi Zhou, Wei Li, and Peter~J Liu.
\newblock {Exploring the Limits of Transfer Learning with a Unified
  Text-to-Text Transformer}.
\newblock {\em The Journal of Machine Learning Research}, 21(1):5485--5551,
  2020.

\bibitem{li2022blip}
Junnan Li, Dongxu Li, Caiming Xiong, and Steven Hoi.
\newblock {BLIP: Bootstrapping Language-Image Pre-training for Unified
  Vision-Language Understanding and Generation}.
\newblock In {\em International Conference on Machine Learning}, pages
  12888--12900. PMLR, 2022.

\bibitem{grigorescu2020survey}
Sorin Grigorescu, Bogdan Trasnea, Tiberiu Cocias, and Gigel Macesanu.
\newblock A survey of deep learning techniques for autonomous driving.
\newblock {\em Journal of Field Robotics}, 37(3):362--386, 2020.

\bibitem{qi2017pointnet}
Charles~R Qi, Hao Su, Kaichun Mo, and Leonidas~J Guibas.
\newblock Pointnet: Deep learning on point sets for 3d classification and
  segmentation.
\newblock In {\em Proceedings of the IEEE conference on computer vision and
  pattern recognition}, pages 652--660, 2017.

\bibitem{zhou2018voxelnet}
Yin Zhou and Oncel Tuzel.
\newblock Voxelnet: End-to-end learning for point cloud based 3d object
  detection.
\newblock In {\em Proceedings of the IEEE conference on computer vision and
  pattern recognition}, pages 4490--4499, 2018.

\bibitem{touvron2023llama}
Hugo Touvron, Thibaut Lavril, Gautier Izacard, Xavier Martinet, Marie-Anne
  Lachaux, Timoth{\'e}e Lacroix, Baptiste Rozi{\`e}re, Naman Goyal, Eric
  Hambro, Faisal Azhar, et~al.
\newblock {LLaMa: Open and Efficient Foundation Language Models}.
\newblock {\em arXiv preprint arXiv:2302.13971}, 2023.

\bibitem{radford2019language}
Alec Radford, Jeffrey Wu, Rewon Child, David Luan, Dario Amodei, Ilya
  Sutskever, et~al.
\newblock {Language models are unsupervised multitask learners}.
\newblock {\em OpenAI blog}, 1(8):9, 2019.

\bibitem{brown2020language}
Tom Brown, Benjamin Mann, Nick Ryder, Melanie Subbiah, Jared~D Kaplan, Prafulla
  Dhariwal, Arvind Neelakantan, Pranav Shyam, Girish Sastry, Amanda Askell,
  et~al.
\newblock {Language models are few-shot learners}.
\newblock {\em Advances in neural information processing systems},
  33:1877--1901, 2020.

\bibitem{wang2022lightseq2}
Xiaohui Wang, Yang Wei, Ying Xiong, Guyue Huang, Xian Qian, Yufei Ding,
  Mingxuan Wang, and Lei Li.
\newblock Lightseq2: Accelerated training for transformer-based models on gpus.
\newblock In {\em SC22: International Conference for High Performance
  Computing, Networking, Storage and Analysis}, pages 1--14. IEEE, 2022.

\bibitem{choquette2022nvidia}
Jack Choquette.
\newblock Nvidia hopper gpu: Scaling performance.
\newblock In {\em 2022 IEEE Hot Chips 34 Symposium (HCS)}, pages 1--46. IEEE
  Computer Society, 2022.

\bibitem{li2020pytorch}
Shen Li, Yanli Zhao, Rohan Varma, Omkar Salpekar, Pieter Noordhuis, Teng Li,
  Adam Paszke, Jeff Smith, Brian Vaughan, Pritam Damania, et~al.
\newblock Pytorch distributed: Experiences on accelerating data parallel
  training.
\newblock {\em arXiv preprint arXiv:2006.15704}, 2020.

\bibitem{huang2019gpipe}
Yanping Huang, Youlong Cheng, Ankur Bapna, Orhan Firat, Dehao Chen, Mia Chen,
  HyoukJoong Lee, Jiquan Ngiam, Quoc~V Le, Yonghui Wu, et~al.
\newblock Gpipe: {E}fficient training of giant neural networks using pipeline
  parallelism.
\newblock {\em Advances in neural information processing systems}, 32, 2019.

\bibitem{shoeybi2019megatron}
Mohammad Shoeybi, Mostofa Patwary, Raul Puri, Patrick LeGresley, Jared Casper,
  and Bryan Catanzaro.
\newblock {Megatron-LM: Training multi-billion parameter language models using
  model parallelism}.
\newblock {\em arXiv preprint arXiv:1909.08053}, 2019.

\bibitem{apple}
Apple.
\newblock {Apple A15 Bionic}, 2021.
\newblock \url{https://en.wikipedia.org/wiki/Apple_A15}.

\bibitem{gao2017tetris}
Mingyu Gao, Jing Pu, Xuan Yang, Mark Horowitz, and Christos Kozyrakis.
\newblock Tetris: {S}calable and efficient neural network acceleration with
  {3D} memory.
\newblock In {\em Proceedings of the Twenty-Second International Conference on
  Architectural Support for Programming Languages and Operating Systems}, pages
  751--764, 2017.

\bibitem{gao2019tangram}
Mingyu Gao, Xuan Yang, Jing Pu, Mark Horowitz, and Christos Kozyrakis.
\newblock Tangram: {O}ptimized coarse-grained dataflow for scalable {NN}
  accelerators.
\newblock In {\em Proceedings of the Twenty-Fourth International Conference on
  Architectural Support for Programming Languages and Operating Systems}, pages
  807--820, 2019.

\bibitem{jouppi2021ten}
Norman~P Jouppi, Doe~Hyun Yoon, Matthew Ashcraft, Mark Gottscho, Thomas~B
  Jablin, George Kurian, James Laudon, Sheng Li, Peter Ma, Xiaoyu Ma, et~al.
\newblock Ten lessons from three generations shaped google’s tpuv4i:
  Industrial product.
\newblock In {\em 2021 ACM/IEEE 48th Annual International Symposium on Computer
  Architecture (ISCA)}, pages 1--14. IEEE, 2021.

\bibitem{shao2019simba}
Yakun~Sophia Shao, Jason Clemons, Rangharajan Venkatesan, Brian Zimmer, Matthew
  Fojtik, Nan Jiang, Ben Keller, Alicia Klinefelter, Nathaniel Pinckney,
  Priyanka Raina, et~al.
\newblock Simba: {S}caling deep-learning inference with multi-chip-module-based
  architecture.
\newblock In {\em Proceedings of the 52nd Annual IEEE/ACM International
  Symposium on Microarchitecture}, pages 14--27, 2019.

\bibitem{wechsler2019spring}
Ofri Wechsler, Michael Behar, and Bharat Daga.
\newblock Spring hill (nnp-i 1000) intel’s data center inference chip.
\newblock In {\em 2019 IEEE Hot Chips 31 Symposium (HCS)}, pages 1--12. IEEE
  Computer Society, 2019.

\bibitem{moon2021evaluating}
Gordon~Euhyun Moon, Hyoukjun Kwon, Geonhwa Jeong, Prasanth Chatarasi,
  Sivasankaran Rajamanickam, and Tushar Krishna.
\newblock Evaluating spatial accelerator architectures with tiled matrix-matrix
  multiplication.
\newblock {\em IEEE Transactions on Parallel and Distributed Systems},
  33(4):1002--1014, 2021.

\bibitem{narayanan2021efficient}
Deepak Narayanan, Mohammad Shoeybi, Jared Casper, Patrick LeGresley, Mostofa
  Patwary, Vijay Korthikanti, Dmitri Vainbrand, Prethvi Kashinkunti, Julie
  Bernauer, Bryan Catanzaro, et~al.
\newblock Efficient large-scale language model training on gpu clusters using
  megatron-lm.
\newblock In {\em Proceedings of the International Conference for High
  Performance Computing, Networking, Storage and Analysis}, pages 1--15, 2021.

\bibitem{shazeer2018mesh}
Noam Shazeer, Youlong Cheng, Niki Parmar, Dustin Tran, Ashish Vaswani, Penporn
  Koanantakool, Peter Hawkins, HyoukJoong Lee, Mingsheng Hong, Cliff Young,
  et~al.
\newblock Mesh-tensorflow: Deep learning for supercomputers.
\newblock {\em Advances in neural information processing systems}, 31, 2018.

\bibitem{narayanan2019pipedream}
Deepak Narayanan, Aaron Harlap, Amar Phanishayee, Vivek Seshadri, Nikhil~R
  Devanur, Gregory~R Ganger, Phillip~B Gibbons, and Matei Zaharia.
\newblock {PipeDream: Generalized pipeline parallelism for DNN training}.
\newblock In {\em Proceedings of the 27th ACM Symposium on Operating Systems
  Principles}, pages 1--15, 2019.

\bibitem{kosson2021pipelined}
Atli Kosson, Vitaliy Chiley, Abhinav Venigalla, Joel Hestness, and Urs Koster.
\newblock Pipelined backpropagation at scale: training large models without
  batches.
\newblock {\em Proceedings of Machine Learning and Systems}, 3:479--501, 2021.

\bibitem{rashidi2022themis}
Saeed Rashidi, William Won, Sudarshan Srinivasan, Srinivas Sridharan, and
  Tushar Krishna.
\newblock {Themis: A network bandwidth-aware collective scheduling policy for
  distributed training of dl models}.
\newblock In {\em Proceedings of the 49th Annual International Symposium on
  Computer Architecture}, pages 581--596, 2022.

\bibitem{klenk2020network}
Benjamin Klenk, Nan Jiang, Greg Thorson, and Larry Dennison.
\newblock An in-network architecture for accelerating shared-memory
  multiprocessor collectives.
\newblock In {\em 2020 ACM/IEEE 47th Annual International Symposium on Computer
  Architecture (ISCA)}, pages 996--1009. IEEE, 2020.

\bibitem{rashidi2020astra}
Saeed Rashidi, Srinivas Sridharan, Sudarshan Srinivasan, and Tushar Krishna.
\newblock {Astra-sim: Enabling sw/hw co-design exploration for distributed dl
  training platforms}.
\newblock In {\em 2020 IEEE International Symposium on Performance Analysis of
  Systems and Software (ISPASS)}, pages 81--92. IEEE, 2020.

\bibitem{moolchandani2023amped}
Diksha Moolchandani, Joyjit Kundu, Frederik Ruelens, Peter Vrancx, Timon
  Evenblij, and Manu Perumkunnil.
\newblock Amped: An analytical model for performance in distributed training of
  transformers.
\newblock In {\em 2023 IEEE International Symposium on Performance Analysis of
  Systems and Software (ISPASS)}, pages 306--315. IEEE, 2023.

\bibitem{geoffrey2021habitat}
X~Yu Geoffrey, Yubo Gao, Pavel Golikov, and Gennady Pekhimenko.
\newblock Habitat: A $\{$Runtime-Based$\}$ computational performance predictor
  for deep neural network training.
\newblock In {\em 2021 USENIX Annual Technical Conference (USENIX ATC 21)},
  pages 503--521, 2021.

\bibitem{james2020ispd}
Michael James, Marvin Tom, Patrick Groeneveld, and Vladimir Kibardin.
\newblock Ispd 2020 physical mapping of neural networks on a wafer-scale deep
  learning accelerator.
\newblock In {\em Proceedings of the 2020 International Symposium on Physical
  Design}, pages 145--149, 2020.

\bibitem{won2023astra}
William Won, Taekyung Heo, Saeed Rashidi, Srinivas Sridharan, Sudarshan
  Srinivasan, and Tushar Krishna.
\newblock Astra-sim2. 0: Modeling hierarchical networks and disaggregated
  systems for large-model training at scale.
\newblock In {\em 2023 IEEE International Symposium on Performance Analysis of
  Systems and Software (ISPASS)}, pages 283--294. IEEE, 2023.

\bibitem{cai2023inter}
Jingwei Cai, Yuchen Wei, Zuotong Wu, Sen Peng, and Kaisheng Ma.
\newblock Inter-layer scheduling space definition and exploration for tiled
  accelerators.
\newblock In {\em Proceedings of the 50th Annual International Symposium on
  Computer Architecture}, pages 1--17, 2023.

\bibitem{vasiljevic2021compute}
Jasmina Vasiljevic, Ljubisa Bajic, Davor Capalija, Stanislav Sokorac, Dragoljub
  Ignjatovic, Lejla Bajic, Milos Trajkovic, Ivan Hamer, Ivan Matosevic,
  Aleksandar Cejkov, et~al.
\newblock Compute substrate for software 2.0.
\newblock {\em IEEE micro}, 41(2):50--55, 2021.

\bibitem{CS}
{Stewart Hall, Rob Schreiber, Sean Lie, Cerebras Systems, Inc.}
\newblock Cs weight streaming white paper.
\newblock
  \url{https://8968533.fs1.hubspotusercontent-na1.net/hubfs/8968533/Virtual
  Booth Docs/CS Weight Streaming White Paper.pdf}, 2023.

\bibitem{ignjatovic2022wormhole}
Drago Ignjatovi{\'c}, Daniel~W Bailey, and Ljubisa Baji{\'c}.
\newblock The wormhole ai training processor.
\newblock In {\em 2022 IEEE International Solid-State Circuits Conference
  (ISSCC)}, volume~65, pages 356--358. IEEE, 2022.

\bibitem{A100}
{Nvidia}.
\newblock Nvidia a100 tensor core gpu architecture.
\newblock
  \url{https://images.nvidia.com/aem-dam/en-zz/Solutions/data-center/nvidia-ampere-architecture-whitepaper.pdf.pdf},
  2017.

\bibitem{dojo}
Emil Talpes, Debjit~Das Sarma, Doug Williams, Sahil Arora, Thomas Kunjan,
  Benjamin Floering, Ankit Jalote, Christopher Hsiong, Chandrasekhar Poorna,
  Vaidehi Samant, John Sicilia, Anantha~Kumar Nivarti, Raghuvir Ramachandran,
  Tim Fischer, Ben Herzberg, Bill McGee, Ganesh Venkataramanan, and Pete Banon.
\newblock The microarchitecture of dojo, tesla’s exa-scale computer.
\newblock {\em IEEE Micro}, 43(3):31--39, 2023.

\bibitem{cerebras}
S.~Lie.
\newblock Cerebras architecture deep dive: First look inside the hw/sw
  co-design for deep learning : Cerebras systems.
\newblock In {\em 2022 IEEE Hot Chips 34 Symposium (HCS)}, pages 1--34, Los
  Alamitos, CA, USA, aug 2022. IEEE Computer Society.

\bibitem{samajdar2018scale}
Ananda Samajdar, Yuhao Zhu, Paul Whatmough, Matthew Mattina, and Tushar
  Krishna.
\newblock Scale-sim: Systolic cnn accelerator simulator.
\newblock {\em arXiv preprint arXiv:1811.02883}, 2018.

\bibitem{opt}
{PyTorch }.
\newblock Torch.optim.
\newblock \url{https://pytorch.org/docs/stable/optim.html}, 2023.

\bibitem{rajbhandari2020zero}
Samyam Rajbhandari, Jeff Rasley, Olatunji Ruwase, and Yuxiong He.
\newblock Zero: Memory optimizations toward training trillion parameter models.
\newblock In {\em SC20: International Conference for High Performance
  Computing, Networking, Storage and Analysis}, pages 1--16. IEEE, 2020.

\bibitem{simpy}
{Klaus G. Müller and Tony Vignaux}.
\newblock Simpy-discrete event simulation for python.
\newblock \url{https://simpy.readthedocs.io/en/latest/}, 2023.

\bibitem{gwennap2020tenstorrent}
Linley Gwennap.
\newblock Tenstorrent scales ai performance: New multicore architecture leads
  in data-center power efficiency, 2020.

\bibitem{Zhuang2022OnOT}
Yonghao Zhuang, Hexu Zhao, Lianmin Zheng, Zhuohan Li, Eric~P. Xing, Qirong Ho,
  Joseph~E. Gonzalez, Ion Stoica, and Haotong Zhang.
\newblock On optimizing the communication of model parallelism.
\newblock {\em ArXiv}, abs/2211.05322, 2022.

\bibitem{liu2022autopipe}
Weijie Liu, Zhiquan Lai, Shengwei Li, Yabo Duan, Keshi Ge, and Dongsheng Li.
\newblock Autopipe: A fast pipeline parallelism approach with balanced
  partitioning and micro-batch slicing.
\newblock In {\em 2022 IEEE International Conference on Cluster Computing
  (CLUSTER)}, pages 301--312. IEEE, 2022.

\end{thebibliography}
\end{document}